\documentclass[final,3p,times,twocolumn]{elsarticle} 

\usepackage{orcidlink}

\usepackage{amssymb}
\usepackage{amsmath}

\usepackage{tikz}
\usetikzlibrary{positioning,arrows.meta,calc,fit}

\usepackage{booktabs}
\usepackage{multirow}

\usepackage{siunitx} 
\usepackage{subcaption}

\usepackage{lineno}

\usepackage{xurl}
\usepackage{hyperref}

\usepackage[table]{xcolor}
\usepackage[normalem]{ulem}
\newcommand{\added}[1]{#1}

\usepackage{colortbl}

\definecolor{good}{RGB}{217,240,211}   
\definecolor{mid}{RGB}{255,243,205}    
\definecolor{bad}{RGB}{248,215,218}    
\definecolor{grey}{HTML}{A6A6A6}       
\colorlet{squrow}{grey!20}             

\journal{Applied Energy}

\begin{document}

\begin{frontmatter}



\title{Surrogate Modeling of Interconnector Flows: A Machine Learning Alternative to Full-Scale Power System Simulations with Application to Cross-Border Electricity Exchange\tnoteref{aam}} 

\tnotetext[aam]{%
This is the accepted manuscript version of an article published in
\textit{Applied Energy}.
The final published version is available at
\href{https://doi.org/10.1016/j.apenergy.2026.128946}
{https://doi.org/10.1016/j.apenergy.2026.128946}.
\textcopyright~2026. This manuscript version is made available under the
\href{https://creativecommons.org/licenses/by-nc-nd/4.0/}
{CC BY-NC-ND 4.0 license}.}


\author[inst1,inst2]{Robert Gaugl\orcidlink{0000-0003-4112-4483}\corref{cor1}}
\ead{robert.gaugl@tugraz.at}

\author[inst3]{Eloy Insunza\orcidlink{0009-0005-9347-4559}}
\ead{einsunza@comillas.edu}

\author[inst3]{José Portela\orcidlink{0000-0002-7839-8982}}
\ead{jportela@comillas.edu}

\author[inst1,inst2]{Sonja Wogrin\orcidlink{0000-0002-3889-7197}}
\ead{wogrin@tugraz.at}

\cortext[cor1]{Corresponding author.}

\affiliation[inst1]{organization={Institute of Electricity Economics and Energy Innovation (IEE), Graz University of Technology},
            addressline={Inffeldgasse 18}, 
            city={Graz},
            postcode={8010}, 
            country={Austria}}

\affiliation[inst2]{organization={Research Center ENERGETIC, Graz University of Technology},
            addressline={Rechbauerstraße 12}, 
            city={Graz},
            postcode={8010}, 
            country={Austria}}

\affiliation[inst3]{organization={Institute for Research in Technology (IIT), ICAI School of Engineering, Universidad Pontificia Comillas},
            addressline={Alberto Aguilera 23}, 
            city={Madrid},
            postcode={28015}, 
            state={Comunidad de Madrid},
            country={Spain}}

\begin{abstract}
Cross-border electricity exchanges are crucial for operating and planning highly renewable power systems. Many studies reduce spatial granularity to keep models tractable and prescribe cross-border exchanges exogenously, often by reusing historical import/export time series. Such assumptions become inconsistent as renewable penetration changes the magnitude and timing of flows. \added{The core contribution of this work is a machine-learning (ML) surrogate framework that maps nodal time-series data (e.g., hourly demand and renewable generation) to synthetic interconnector-level flow time series. The resulting flows serve as fixed boundary conditions in reduced power system optimization models (PSOMs). This addresses studies focused on a subset of an interconnected system, for which repeatedly solving the full model across multiple climate years or scenarios is computationally expensive.} We demonstrate the framework on a pan-European single-node-per-country DC optimal power flow model. We benchmark $k$-nearest neighbors (KNN), ridge regression, random forests, Extra Trees, gradient boosting, XGBoost, and feedforward neural-network (SQU) surrogates. As an additional contribution, we introduce a novel custom neural-network loss that softly penalizes physically implausible flow patterns. This soft penalty is not a guarantee of full power-flow feasibility. Across benchmarks, the SQU variants provide the most consistent out-of-sample and downstream optimization performance while substantially outperforming scaled historical profiles. The ML-generated flows closely reproduce country-level results of the full European model while greatly reducing computation time (up to $\sim$500$\times$). The proposed framework enables scenario-consistent reduced PSOM studies across climate years after a single full-model run, eliminating repeated pan-European simulations and supporting more robust climate-year analyses under fixed system assumptions.

\end{abstract}




\end{frontmatter}



\section{Introduction}
\label{sec:Introduction}

\subsection{Motivation}
\label{sec:Motivation}

The European power system is characterized by a high degree of interconnectivity among national electricity grids, making cross-border power flows a critical element of both system operation and long-term planning. In future power systems with high shares of variable renewable energy sources (VRES), cross-border electricity trading will become even more essential for balancing fluctuations in VRES generation across regions. Recognizing this, the European Union (EU) has introduced regulatory measures mandating that at least \SI{70}{\percent} of cross-border transmission capacity be made available for electricity trading, as outlined in the Clean Energy for All Europeans package \citep{european_union_regulation_2019}.

Power system optimization models (PSOMs) are widely used to analyze such systems and guide infrastructure and policy decisions. However, PSOMs face inherent trade-offs due to computational limitations, particularly across three key dimensions: (1) temporal resolution, (2) technical detail, and (3) spatial scope. Even with access to modern high-performance computing, it remains challenging to fully capture all three dimensions simultaneously. Consequently, many studies restrict the spatial domain to a single country or region to preserve detail in other aspects. Yet, electricity imports and exports are crucial to capture cross-border dynamics. Therefore, broader spatial modeling is often needed to ensure system-wide interactions are adequately represented, especially those involving international electricity trade.

To reduce the computational burden of large-scale spatial PSOMs, cross-border exchanges are frequently treated exogenously, for example by prescribing historical or otherwise fixed import and export time series. This type of simplification is common in geographically restricted modeling studies, where cross-border trade is represented in a stylized way to retain tractability, but can materially affect results as shown by \citet{mertens_representing_2020}. However, as the European power system transitions toward a higher share of VRES, the patterns of cross-border electricity flows are already undergoing significant changes. Imports and exports increasingly reflect fluctuating renewable availability rather than predictable demand-driven patterns \citep{crozier_effect_2022, unger_effect_2018}. As a result, scaling historical time series---a common practical workaround when full-system simulations are computationally prohibitive---can lead to misleading assumptions and inaccurate model outcomes if these structural changes are ignored.

\added{This modeling problem occurs frequently in applied energy-system research: the question of interest may concern only one country, region, or a detailed lower-level grid, while that subsystem remains influenced by the surrounding interconnected system. Explicitly co-optimizing a detailed regional model together with the full European system for every climate year or sensitivity can substantially increase computational requirements. The proposed framework therefore targets energy-system and scenario analysis rather than real-time system operation, security assessment, or detailed grid-feasibility studies. Applications that require guaranteed AC or DC power-flow feasibility, security constraints, or operational set points still require the corresponding physics-based network model.}

\subsection{Literature Review}
\label{sec:LiteratureReview}

Over the past decade, machine learning (ML) techniques have become central to forecasting tasks in the energy sector. Among the most established applications are electricity demand and market price forecasting, where numerous studies in high-impact journals have demonstrated the reliability and accuracy of ML models. For instance, deep learning and ensemble methods have shown strong performance in predicting short-term electricity demand \citep{bedi_deep_2019, gajowniczek_two-stage_2017, nowotarski_recent_2018, lago_forecasting_2021, bille_forecasting_2023}, while similar approaches have been widely applied for electricity price forecasting \citep{tschora_electricity_2022, nitsch_applying_2024, hong_probabilistic_2016, eren_comprehensive_2024, biswal_review_2024}.

More recently, ML has gained traction as a surrogate for computationally intensive power system optimization tasks, including optimal power flow (OPF), unit commitment (UC), and economic dispatch (ED), where repeated solves for uncertainty analysis or scenario screening can be prohibitive. In the OPF context, \citet{mohammadi_surrogate_2024} review both analytical surrogates (e.g., linearized or reduced-order models) and learned surrogates trained on large sets of OPF solutions, highlighting that data-driven approaches can yield substantial runtime reductions while retaining high accuracy. Existing ML-based OPF surrogates commonly predict dispatch decisions, identify active constraints, approximate optimal solutions, or provide warm starts for the original optimization problem \citep{khaloie_review_2025}. For UC and ED, ML models are similarly used to approximate commitment decisions or accelerate optimization, although their performance must be assessed against strong heuristic baselines \citep{pineda_is_2022}. Recent work nonetheless reports encouraging results, including support vector machine-based warm-start strategies \citep{pourahmadi_unit_2025} and reinforcement-learning-assisted methods embedded in decomposition frameworks \citep{zhu_reinforcement_2025, huang_intelligent_2023}.


\added{The core contribution of the present work lies in a specific use of surrogate modeling: rather than predicting the complete solution of an OPF or accelerating the solution of the original optimization problem, we predict interconnector-level cross-border flows generated by a full PSOM and subsequently use these learned flows as fixed boundary conditions in geographically reduced PSOMs. This addresses a recurring problem in power-system studies in which the research question concerns only a country or region, while the surrounding interconnected system still materially influences the modeled region through imports and exports. Explicitly representing the complete surrounding system for every climate year or scenario can substantially increase computational requirements, whereas omitting it neglects relevant external-system interactions.}

\added{Methodologically, this core contribution is implemented within a predict-then-optimize structure} \citep{elmachtoub_smart_2022}: interconnector flows are first predicted and are subsequently imposed as fixed parameters in a reduced PSOM. Unlike decision-focused learning, where the predictive model is trained according to the quality of the resulting downstream decision \citep{wilder_melding_2019}, the surrogate developed here is trained to reproduce boundary flows generated by the full PSOM. Predictive accuracy and downstream optimization performance are therefore evaluated separately.

Beyond the application-oriented development of ML surrogates for OPF, another important methodological research direction is physics-informed machine learning. Rather than relying solely on data-driven training objectives, physics-informed approaches embed physical knowledge directly into the learning process through governing equations or physically motivated constraints \citep{raissi_physics-informed_2019,karniadakis_physics-informed_2021}. \added{The feasibility-aware custom loss proposed in this work is inspired by the same general philosophy by incorporating domain-specific operational knowledge directly into the learning objective. Rather than enforcing the full DC-OPF equations or power-flow physics, it introduces a soft penalty for net-export patterns that are inconsistent with an approximate domestic supply bound and may induce artificial ENS when imposed in a reduced PSOM. The penalty is not a projection onto the DC-OPF feasible set and does not provide a formal feasibility guarantee.} Extending this idea toward richer physics-informed formulations that incorporate additional power-system constraints represents a promising direction for future research.


In parallel, the power-system modeling literature documents that cross-border exchanges are often simplified exogenously to retain tractability. For example, \citet{frey_tackling_2024} employ historical import/export profiles within a coupled modeling framework and explicitly note that such assumptions can become problematic as the energy transition alters both the magnitude and temporal structure of cross-border exchanges. Similarly, \citet{aravena_renewable_2017} model exchanges between the CWE region and non-CWE countries as fixed flows in a zonal market context. \added{\citet{mertens_representing_2020} directly demonstrate the importance of this modeling problem for long-term optimization models with limited geographical scope: the treatment of cross-border electricity trade can affect technology choices and welfare estimates, motivating explicit representations of the external system. Their work establishes the modeling problem addressed here, but uses model-based representations of foreign electricity trade rather than learning interconnector-level flow time series from a full-system PSOM.}

\added{The comparison with adjacent ML-based optimization literature further sharpens the gap. Representative OPF-surrogate studies learn variables of the original optimization problem and seek to approximate or accelerate the original OPF, including feasibility-aware approaches that incorporate engineering constraints \citep{fioretto_predicting_2020,zamzam_learning_2020,donti_dc3_2021}. These methods address an important but different task from transferring the effect of an external interconnected system to a geographically reduced PSOM. Thus, the individual elements of the proposed framework have precedents in different research streams, but our literature review did not identify previous work that combines them in the same way: full-system PSOM results provide interconnector-flow training targets, the learned interconnector flows become fixed boundary conditions of a geographically reduced PSOM, and their quality is assessed through downstream optimization outcomes. This combination constitutes the core contribution of the proposed framework.}

\begin{table*}[t]
\centering
\caption{\added{Comparison of the proposed framework with representative literature most closely related to its core contribution. A checkmark indicates that the corresponding element is an explicit part of the cited approach.}}
\label{tab:literature_comparison}
\small
\resizebox{\textwidth}{!}{%
\begin{tabular}{lcccccc}
\toprule
Study & ML used & Cross-border flow prediction & Full-PSOM flow targets & External-system representation & Learned flows as reduced-PSOM boundaries & Downstream reduced-PSOM validation \\
\midrule
\citet{mertens_representing_2020} & -- & -- & -- & \checkmark & -- & \checkmark \\
\citet{frey_tackling_2024} & -- & -- & -- & \checkmark & -- & \checkmark \\
\citet{aravena_renewable_2017} & -- & -- & -- & \checkmark & -- & -- \\
\citet{kledzik_predicting_2024} & \checkmark & \checkmark & -- & -- & -- & -- \\
\citet{fioretto_predicting_2020} & \checkmark & -- & --$^{a}$ & -- & -- & -- \\
\citet{zamzam_learning_2020} & \checkmark & -- & --$^{a}$ & -- & -- & -- \\
\citet{donti_dc3_2021} & \checkmark & -- & --$^{a}$ & -- & -- & -- \\
\textbf{This work} & \checkmark & \checkmark & \checkmark & \checkmark & \checkmark & \checkmark \\
\bottomrule
\end{tabular}}
\begin{flushleft}
\footnotesize
\added{$^{a}$These studies learn variables associated with the original constrained optimization problem rather than interconnector-flow targets that are transferred as boundary conditions to a geographically reduced PSOM.}
\end{flushleft}
\end{table*}

\added{Table~\ref{tab:literature_comparison} makes the specific literature gap explicit. We do not claim that ML-based OPF surrogates, ML-based cross-border-flow prediction, or geographically reduced energy-system models are individually new. The core contribution is their connection in a reusable boundary-flow surrogate framework: interconnector-level flows generated by a full PSOM are learned for subsequent use as boundary conditions in a geographically reduced PSOM, and the resulting reduced-model costs, dispatch, storage operation, trade, and ENS are compared with the corresponding full-system outcomes.}

\subsection{Original Contributions and Proposed ML Models}
\label{sec:POriginalContributionAndroposedMLModels}


\added{Given the background outlined above, the core contribution is a framework for learning cross-border boundary information from a full European PSOM and transferring this information to geographically reduced PSOM studies. The framework is intended for prospective energy-system planning and scenario analysis in which demand and renewable deployment are specified exogenously and weather variability is represented through different climate years. It is not proposed as a replacement for detailed power-flow, security, or real-time operational analysis.}

Therefore, this paper makes the following research contributions:
\begin{itemize}
    \item \added{\textbf{Core contribution:} We propose an ML-based surrogate framework that predicts hourly interconnector-level cross-border flows generated by a full PSOM and uses them as fixed boundary conditions in geographically reduced PSOMs. This enables repeated reduced-model studies across different climate years without repeatedly solving the full European PSOM under unchanged structural assumptions.}
    \item \added{We demonstrate the decision relevance of the predicted boundary flows by evaluating not only statistical flow-prediction accuracy but also their effects on downstream reduced-PSOM results, including system costs, generation dispatch, storage operation, cross-border trade, and ENS, relative to the corresponding country-level results of the full European model.}
    \item \added{\textbf{Additional methodological contribution:} We introduce a feasibility-aware neural-network loss that applies a soft penalty to net-export patterns that are inconsistent with an approximate domestic supply bound when the predicted interconnector flows are intended for subsequent use in a reduced PSOM. The loss is explicitly treated as a soft plausibility mechanism, not as a guarantee of complete power-flow feasibility.}
    \item \added{We benchmark the proposed approach against KNN, ridge regression, random forests, Extra Trees, gradient boosting, XGBoost, multiple neural-network variants, and historical/scaled import--export profiles across Austria, Germany, and Spain and out-of-sample climate years.}
    \item We generate training targets with the pan-European LEGO DC-OPF (single node per country) under TYNDP~2024 \emph{National Trends} 2030 assumptions, train on climate year (CY) 2009, and evaluate both predictive accuracy and downstream single-country PSOM impacts on the out-of-sample CYs~1995 and~2008 to test robustness across weather realizations.
\end{itemize}

The remainder of the paper is structured as follows. Section~\ref{sec:Methodology} presents the generic surrogate workflow, including the PSOM formulation used to generate interconnector-flow targets, the ML problem definition, and the KNN, ridge regression, random forests, Extra Trees, gradient boosting, XGBoost and SQU surrogate model variants (including the feasibility-aware custom-loss formulation). Section~\ref{sec:CaseStudy} then instantiates this methodology for our empirical setting by describing the simplified European single-node-per-country DC-OPF setup, the scenario data and climate years used for training and out-of-sample evaluation, and the selected case-study countries. Section~\ref{sec:Results} reports predictive performance and quantifies how surrogate-generated boundary flows affect reduced single-country PSOM outcomes relative to the corresponding country components of the full European benchmark. Finally, Sections~\ref{sec:Discussion} and~\ref{sec:ConclusionOutlook} discuss the implications of the findings and outline directions for future work.

\section{Methodology}
\label{sec:Methodology}

\begin{figure*}
\centering
\resizebox{\textwidth}{!}{%
  \includegraphics[width=\textwidth]{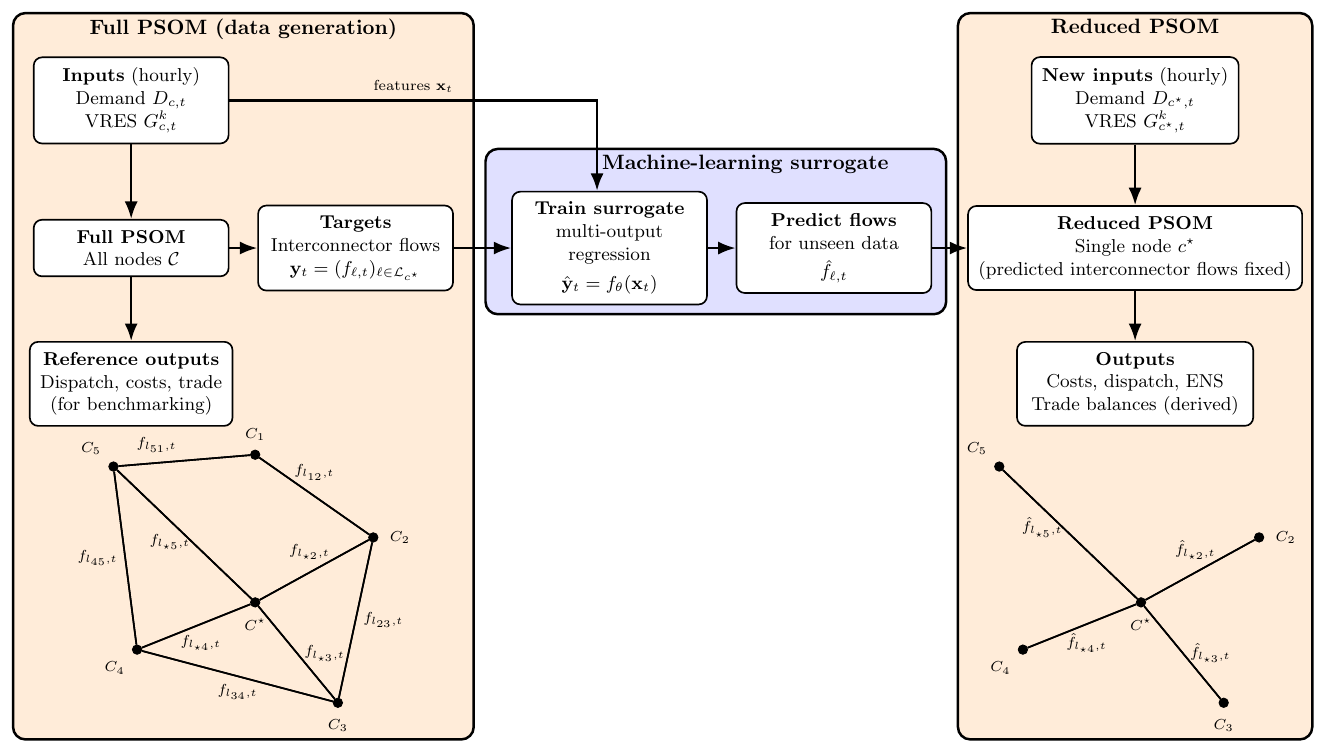}
}
\caption{Workflow of the proposed approach. A full PSOM is run to generate hourly interconnector-flow targets for the node of interest. Full-feature baselines (KNN, ridge regression, random forests, Extra Trees, gradient boosting, and XGBoost) and SQU neural-network variants are trained on demand and VRES inputs and predict flows for unseen data (e.g., a different climate year). The predicted flows are imposed as fixed boundary conditions in a reduced single-node PSOM, and the resulting costs, dispatch, ENS, and trade are compared against the corresponding outcomes from the full PSOM. This workflow enables repeated geographically reduced PSOM studies that retain a model-consistent representation of cross-border interaction while avoiding repeated pan-European optimization runs. \added{The reduced PSOM provides a downstream consistency check; scenarios in which surrogate-generated boundary conditions induce unacceptable behavior can be rerun with the full European PSOM.}}
\label{fig:workflow}
\end{figure*}

We propose a surrogate framework (illustrated in Figure~\ref{fig:workflow}) that maps node-level renewable generation and demand time series to hourly interconnector flows. We use the open-source LEGO PSOM \citep{wogrin_lego_2022} to generate training data for the ML models and to compute the reference results. The surrogates are trained on inputs comprising renewable generation profiles (including PV, wind, run-of-river hydro, and concentrated-solar-power (CSP)) and demand per node, with the target outputs being the interconnector flows produced by the PSOM. \added{As an additional contribution, we develop a feasibility-aware custom loss for the neural-network surrogate. The loss softly penalizes predicted net exports that exceed an approximate domestic supply bound and may otherwise induce artificial energy-not-supplied (ENS) when the flows are imposed in a reduced PSOM. It does not enforce the full DC-OPF feasible region and is not presented as a guarantee of power-flow feasibility.}

We evaluate predictive performance by comparing surrogate-generated interconnector flows to those obtained from the full PSOM. In a second step, the predicted interconnector flows are used as fixed inputs in a simplified PSOM with spatial resolution restricted to the node of interest. This allows assessing decision relevance by comparing deviations in key outputs (objective value, trade balances, and dispatch indicators) between reduced- and full-scale PSOM runs.

\added{The reduced PSOM also serves as a downstream consistency check for the predicted boundary conditions. If a predicted flow profile induces ENS or otherwise yields an unacceptable reduced-model solution, the corresponding climate year or scenario can be identified and evaluated with the full European PSOM instead. The surrogate is therefore intended to reduce the number of computationally expensive full-system runs, rather than to eliminate the full model or provide a formal feasibility guarantee for every predicted scenario.}

To facilitate understanding of the workflow, the remainder of this section is organized as follows. Section~\ref{sec:PSOM} introduces the PSOM formulation used to generate interconnector-flow targets. Section~\ref{sec:Surrogate} then defines the supervised learning problem, including target variables, feature construction, and the surrogate families, along with the feasibility-aware custom-loss variant. It also describes preprocessing, training, and the evaluation metrics. Finally, Section~\ref{sec:SingleCountry} explains how surrogate-predicted flows are imposed as fixed boundary conditions in a reduced PSOM and how decision relevance is assessed against the corresponding full-model benchmark outcomes.

\subsection{Full PSOM}
\label{sec:PSOM}

The open-source LEGO model is used in order to generate training data for the ML models and to compute the reference results. LEGO is formulated as a linear optimization model based on a DC-OPF approximation.

Let $\mathcal{G}$ denote the set of generators. We distinguish between dispatchable thermal generators $\mathcal{G}^{\text{th}}$, variable renewable generators $\mathcal{G}^{\text{vres}}$, and storage units $\mathcal{G}^{\text{st}}$, while $\mathcal{T}$ denotes the set of time steps:
\begin{equation}
\resizebox{0.9\columnwidth}{!}{$
\begin{aligned}
\min\; f_{\text{obj}}
= \sum_{t \in \mathcal{T}} \Biggl\{ 
& \sum_{g \in \mathcal{G}^{\text{th}}}
\Bigl(
  C^{\text{su}}_{g}\,\mathit{su}_{g,t}
+ C^{\text{com}}_{g}\,u_{g,t}
+ C^{\text{var}}_{g}\,p_{g,t}
\Bigr) \\
& + \sum_{r \in \mathcal{G}^{\text{vres}}} C^{\text{var}}_{r}\,p_{r,t}
+ \sum_{s \in \mathcal{G}^{\text{st}}}   C^{\text{var}}_{s}\,p^{\text{dis}}_{s,t} \\
& + P^{\text{CO}_2} \sum_{g \in \mathcal{G}^{\text{th}}}
\varepsilon_{g}
\Bigl(
  \alpha^{\text{su}}_{g}\,\mathit{su}_{g,t}
+ \alpha^{\text{int}}_{g}\,u_{g,t}
+ \alpha^{\text{var}}_{g}\,p_{g,t}
\Bigr)
\Biggr\}.
\end{aligned}
$}
\label{eq:objective}
\end{equation}

The objective function in \eqref{eq:objective} minimizes the total operational costs of all generation units across all nodes and time steps. It consists of two main components. The first component captures the conventional techno\textendash economic operating costs. For dispatchable thermal units $g \in \mathcal{G}^{\text{th}}$, these costs include per time-step startup costs $C^{\text{su}}_{g}\,\mathit{su}_{g,t}$, commitment costs $C^{\text{com}}_{g}\,u_{g,t}$, and generation\textendash dependent variable costs $C^{\text{var}}_{g}\,p_{g,t}$. The binary variables $\mathit{su}_{g,t} \in \{0,1\}$ and $u_{g,t} \in \{0,1\}$ indicate whether thermal unit $g$ starts up or is committed at time $t$. For variable renewable generators $r \in \mathcal{G}^{\text{vres}}$ and storage units $s \in \mathcal{G}^{\text{st}}$, we account for variable operating costs through $C^{\text{var}}_{r}\,p_{r,t}$ and $C^{\text{var}}_{s}\,p^{\text{dis}}_{s,t}$, respectively. The second component represents CO$_2$ emission costs associated with fossil fuel consumption and applies to thermal units only. These costs are computed by multiplying a uniform CO$_2$ price $P^{\text{CO}_2}$ with the emission factor $\varepsilon_{g}$ and the fuel consumption linked to each operating decision: startups $\alpha^{\text{su}}_{g}\,\mathit{su}_{g,t}$, commitment $\alpha^{\text{int}}_{g}\,u_{g,t}$, and energy production $\alpha^{\text{var}}_{g}\,p_{g,t}$.

The main constraints of the model represent standard power system relationships. For each node $c$ and time-step $t$ we impose a nodal power balance constraint:
\begin{equation}
\begin{aligned}
0
= {} D_{c,t}
&- \sum_{g \in \mathcal{G}^{\text{th}}_c} p_{g,t}
- \sum_{r \in \mathcal{G}^{\text{vres}}_c} p_{r,t} \\
& - \sum_{s \in \mathcal{G}^{\text{st}}_c} p^{\text{dis}}_{s,t}
+ \sum_{s \in \mathcal{G}^{\text{st}}_c} p^{\text{ch}}_{s,t}
- \sum_{\ell \in \mathcal{L}_c} f_{\ell,t},
\quad \forall c,t.
\end{aligned}
\label{eq:balance}
\end{equation}

where $D_{c,t}$ is the demand of node $c$ at time $t$, $p_{g,t}$ denotes thermal generation, $p_{r,t}$ denotes variable renewable generation, $p^{\text{ch}}_{s,t}$ and $p^{\text{dis}}_{s,t}$ denote storage charging and discharging power, respectively, $f_{\ell,t}$ denotes the flow on interconnector $\ell$ at time $t$, and $\mathcal{L}_c$ denotes the set of lines from or to node $c$.\footnote{Positive $f_{\ell,t}$ denotes flow into node $c$.} For each unit we impose generation and power limits:

\begin{align}
0 &\le p_{g,t} \le \overline{P}_{g}\, 
&& \forall g \in \mathcal{G}^{\text{th}},\; t, \label{eq:cap_th}\\
0 &\le p_{r,t} \le \overline{P}_{r,t},
&& \forall r \in \mathcal{G}^{\text{vres}},\; t, \label{eq:cap_vres}\\
0 &\le p^{\text{ch}}_{s,t} \le \overline{P}^{\text{ch}}_{s},
&& \forall s \in \mathcal{G}^{\text{st}},\; t, \label{eq:cap_st_ch}\\
0 &\le p^{\text{dis}}_{s,t} \le \overline{P}^{\text{dis}}_{s},
&& \forall s \in \mathcal{G}^{\text{st}},\; t. \label{eq:cap_st_dis}
\end{align}
where $\overline{P}_{g}$ is the installed capacity of thermal unit $g$, and $\overline{P}^{\text{ch}}_{s}$ and  $\overline{P}^{\text{dis}}_{s}$ denote the maximum charging and discharging power of storage unit $s$, respectively. For VRES units, $\overline{P}_{r,t}$ is the time-dependent available power, given by installed capacity times the corresponding availability (capacity-factor) profile. For each interconnector $\ell$ and time-step $t$ we impose line capacity constraints:
\begin{equation}
-\overline{F}_{\ell} \leq f_{\ell,t} \leq \overline{F}_{\ell}, \quad \forall \ell, t,
\end{equation}
where $\overline{F}_{\ell}$ denotes the thermal limit of interconnector $\ell$. The complete LEGO formulation, including the DC-OPF formulation, is given in \citet{wogrin_assessing_2020}. For brevity, we restrict the presentation here to the components required for the surrogate modeling setup.

Running LEGO yields dispatch per time-step as well as the interconnector flows for all nodes.

\subsection{Machine-learning Surrogate}
\label{sec:Surrogate}
For the ML surrogate, we first define the input features and target variables (Section~\ref{sec:Targets}). We then compare six full-feature regression baselines (KNN, ridge regression, random forests, Extra Trees, gradient boosting, and XGBoost) (Section~\ref{sec:tabular_baselines}) with a feedforward sequential neural-network surrogate (SQU) (Section~\ref{sec:SQUModels}). The baselines are implemented using \texttt{scikit-learn}~\citep{pedregosa_scikit-learn_2011} and \texttt{XGBoost}~\citep{contributors_xgboost_2025}, while the neural-network models are implemented in \texttt{TensorFlow/Keras}~\citep{chollet_keras_2015}.

\subsubsection{Target Variables \& Feature Construction}
\label{sec:Targets}

The full LEGO model computes for each interconnector $\ell$ and time-step $t$ the power flow $f_{\ell,t}$. For a node of interest $c^*$, we define $\mathcal{L}_{c^*}$ as the set of all interconnectors that connect $c^*$ with other nodes.

In this work, the machine--learning model is trained to reproduce the \emph{complete vector of flows} on these interconnectors, which we obtain from the full LEGO run. Thus, for each time-step $t$, the target variable is the vector 

\begin{equation}
\mathbf{y}_t
=
\left( f_{\ell,t} \right)_{\ell \in \mathcal{L}_{c^*}},
\qquad t \in \mathcal{T}^{\text{train}},
\label{eq:target_vector}
\end{equation}
which contains one entry for every interconnector linking the node of interest to other nodes.

The supervised target should not be interpreted as a unique physical mapping from demand and renewable-generation profiles to interconnector flows. In a linear OPF with aggregated technologies, identical exogenous inputs may admit multiple cost-equivalent or near-equivalent dispatch solutions, which can imply different net injections and interconnector flows. The target used here is therefore the deterministic solution produced by the specific full PSOM formulation, input data, solver configuration, numerical tolerances, and solver tie-breaking behavior. Consequently, the surrogate learns the mapping induced by this complete modeling and solution pipeline rather than a unique physical law.

The feature vector $\mathbf{x}_t$ is constructed from the time-step profiles per node of electricity demand together with the renewable generation time series, namely CSP, PV, onshore wind, offshore wind, and run-of-river hydropower. In the most general case, the feature vector for time-step $t$ is given by
\begin{equation}
\mathbf{x}_t = \bigl[ \, D_{c,t}, \{ G^{k}_{c,t} \}_{k \in \mathcal{K}}, \, c \in \mathcal{C} \, \bigr],
\end{equation}
where $\mathcal{C}$ denotes the set of nodes and $\mathcal{K}$ denotes the set of renewable technologies (PV, onshore wind, offshore wind, run-of-river, CSP), and $G^{k}_{c,t}$ is the generation at time-step $t$ of technology $k$ at node $c$. This feature set represents the complete information available to the PSOM. We refer to this comprehensive feature representation as \emph{Full}.

All baseline benchmark models use the complete \emph{Full} feature set so that differences among them reflect model-class performance rather than different feature-selection pipelines. To investigate the trade-off between neural-network complexity and predictive performance, we additionally consider a reduced SQU feature set. This set, denoted as \emph{Selected}, contains the most influential demand and renewable-generation variables identified through the neural-network sensitivity analysis described in Section~\ref{sec:SQUModels}.

We aim to learn a mapping
\begin{equation}
\mathbf{\hat{y}}_t = f_\theta(\mathbf{x}_t),
\end{equation}
where $\theta$ denotes the ML model parameters (e.g., the weights and biases of the neural network in the SQU models; for the KNN baseline, $\theta$ represents the chosen hyperparameters together with the stored training instances used for neighbor lookup). The learned surrogate $f_\theta(\mathbf{x}_t)$ approximates the interconnector flows generated by the full LEGO model:
\begin{equation}
f_\theta(\mathbf{x}_t) \approx \mathbf{y}_t, \quad \forall t \in \mathcal{T}^{\text{train}}.
\end{equation}

\subsubsection{Full-feature Baselines}
\label{sec:tabular_baselines}

The set of full-feature baseline models comprises KNN, ridge regression, random forests, Extra Trees, gradient boosting, and XGBoost. All baseline models are trained using the identical full input feature set to isolate differences attributable to the regression model itself rather than the feature-selection procedure. Their hyperparameter settings are summarized in Table~\ref{tab:model_configurations}.

As a non-parametric baseline, we employ KNN regression. For each time-step $t$, the model predicts the interconnector flow value $\mathbf{\hat{y}}_t$ as the average of the observed interconnector flow values associated with the $k$ most similar historical system states. Similarity is measured using the Euclidean distance in the standardized feature space defined by $\mathbf{x}_t$. Ridge regression provides a regularized linear baseline by estimating a linear mapping between the input features and interconnector flows while penalizing large regression coefficients. Random forests represent a bagged ensemble of decision trees that reduces prediction variance through bootstrap aggregation. Extra Trees extend this ensemble approach by introducing additional randomness in the feature and split selection during tree construction. Gradient boosting builds an additive ensemble of decision trees sequentially, with each tree correcting the residual errors of the previous ensemble. XGBoost follows the same boosting principle while incorporating second-order optimization, regularization, and efficient tree construction to improve predictive performance and computational efficiency.

All models predict the complete vector of interconnector flows in a multi-output setting. Models that do not natively support multi-output regression are fitted using one regressor per interconnector output. No feature selection or hyperparameter optimization is applied to these baselines, providing a transparent comparison of different regression model classes under the same information set.

\subsubsection{Neural Network Model (SQU)}
\label{sec:SQUModels}

The SQU model is a fully connected feedforward sequential neural network with multiple dense hidden layers and rectified linear unit (ReLU) activation functions.

The baseline configuration, \texttt{SQU\_FULL}, is trained using the standard mean squared error (MSE) loss:

\begin{equation}
\mathcal{L}_{\text{MSE}}(\theta)
= \frac{1}{|\mathcal{T}^{\text{train}}|} 
\sum_{t \in \mathcal{T}^{\text{train}}} 
\left\lVert \mathbf{y}_t - f_\theta(\mathbf{x}_t) \right\rVert_2^2.
\end{equation}

In addition to \texttt{SQU\_FULL}, we evaluate a reduced-feature variant (\texttt{SQU\_SELECTED}) and a hyperparameter-tuned variant (\texttt{SQU\_OPTIMIZED}).

For the neural networks, feature importance is derived from a sensitivity analysis based on the Jacobian of the trained model using the neuralsens python package \citep{pizarroso_neuralsens_2022}. Specifically, we compute the partial derivatives of each output with respect to each standardized input feature across the training set. The standard deviation of these sensitivities serves as an importance measure, reflecting how strongly variations in a given feature influence predicted interconnector flows. For each output, the most influential features are selected, and their union defines the \texttt{SQU\_SELECTED} feature set. Specifically, feature importance is evaluated separately for each interconnector output. The twenty highest-ranked features for each output are combined, and duplicate entries are removed to form a single reduced feature set used for training all outputs simultaneously. The threshold of twenty features per output is applied consistently across all case-study countries as a pragmatic compromise between retaining output-specific information and substantially reducing the dimensionality of the full 249-feature input space.

In addition, we formulate a novel feasibility-aware loss function (\texttt{SQU\_CUSTOM\_LOSS}) \added{designed to discourage interconnector-flow predictions that are inconsistent with an approximate domestic supply bound}. Unlike generic regularization terms, the proposed loss is specifically tailored to the surrogate prediction of interconnector boundary conditions for geographically reduced PSOMs. Its purpose is to improve the operational consistency of the predicted exchanges when they are imposed as fixed boundary conditions in reduced optimization models. In addition to minimizing the squared error of the interconnector-flow vector, the loss includes a penalty for excessive predicted net exports. In particular, it discourages situations in which the implied exports exceeds the nodes maximum available generation capacity plus inflows to the node minus demand, which would otherwise lead to artificial ENS events in the reduced PSOM. Rather than enforcing the complete DC-OPF equations, the proposed loss incorporates a physically motivated operational constraint directly into the learning objective. This follows the general philosophy of physics-informed machine learning, where prior physical knowledge is embedded into the optimization process through additional loss terms.

To operationalize this idea, we precompute a time-step-specific residual bound $R^{\max}_t$ (in MW) from exogenous demand and capacity information for $c^*$. Physically, $R^{\max}_t$ represents an upper bound on the net export that node $c^*$ can sustain in hour~$t$ given demand and maximum available nodal supply. If the predicted net export exceeds this bound, the implied exchange profile cannot be balanced by the reduced single-node PSOM without creating artificial ENS:
\begin{equation}
\begin{split}
R^{\max}_t
=
- D_{c^*,t}
+ \sum_{k \in \mathcal{K}} G^{k}_{c^*,t}
+ \overline{P}^{\text{thermal}}_{c^*} \\
\qquad
+ \overline{P}^{\text{storage}}_{c^*}
+ \overline{P}^{\text{OtherRES}}_{c^*,t},
\end{split}
\label{eq:residual_max}
\end{equation}
where $D_{c^*,t}$ is demand, $G^{k}_{c^*,t}$ denotes non-dispatchable renewable generation time series included in the input dataset, $\overline{P}^{\text{thermal}}_{c^*}$ is the aggregate maximum thermal production capacity, $\overline{P}^{\text{storage}}_{c^*}$ is the aggregate maximum storage discharge capacity, and $\overline{P}^{\text{OtherRES}}_{c^*,t}$ is the production of the \textit{OtherRES} category.

We orient interconnector flows such that $\tilde{\hat y}_{\ell,t}$ is negative for inflow to $c^*$ and positive for outflow.\footnote{This orientation is applied to ensure a consistent sign convention across interconnectors, since interconnectors may be defined in either direction (from $c^*$ to a neighbor or vice versa). By multiplying each interconnector flow by an interface-specific sign factor, we enforce that inflows to $c^*$ are always negative and outflows are always positive, independent of the original interconnector definition.} Accordingly, the net export is given by $\sum_{\ell \in \mathcal{L}_{c^*}} \tilde{\hat y}_{\ell,t}$ (positive for net exports, negative for net imports). The feasibility penalty activates when this predicted net export exceeds the residual bound $R_t^{\max}$.
\begin{equation}
\begin{split}
\mathcal{L}_{\text{custom}}(\theta)
&= \frac{1}{|\mathcal{T}^{\text{train}}|}
\sum_{t \in \mathcal{T}^{\text{train}}}
\Biggl[
\frac{1}{|\mathcal{L}_{c^*}|}
\left\lVert
\tilde{\mathbf{y}}_t - \tilde{\mathbf{\hat y}}_t
\right\rVert_2^2 \\
&\qquad
+\;
\lambda\,
\max\Biggl\{0,\;
\sum_{\ell\in\mathcal{L}_{c^*}} \tilde{\hat y}_{\ell,t}
- R^{\max}_t
\Biggr\}
\Biggr].
\end{split}
\label{eq:customLoss}
\end{equation}

where $\lambda \ge 0$ controls the penalty strength.

\subsubsection{Training Procedure and Evaluation Metrics}
\label{sec:Training}

All input features are standardized using the \texttt{StandardScaler} implementation from the \texttt{scikit-learn} library. The scaling parameters (mean and standard deviation) are estimated exclusively from the training and subsequently applied unchanged to all validation and out-of-sample datasets. For the neural networks, early stopping based on validation loss is employed to prevent overfitting.

Predictive performance is evaluated using the normalized mean absolute error (NMAE) and the coefficient of determination $R^2$:
\begin{align}
\text{NMAE} &= \frac{\sum_{t=1}^N |y_t - \hat{y}_t|}
{\sum_{t=1}^N |y_t|}, \\
R^2 &= 1 - \frac{\sum_{t=1}^N (y_t - \hat{y}_t)^2}
{\sum_{t=1}^N (y_t - \bar{y})^2},
\end{align}

where $y_t$ denotes the real interconnector flow obtained from the full PSOM, $\hat{y}_t$ the corresponding ML prediction, $N$ the number of evaluated time-steps, and $\bar{y}$ the mean of the reference interconnector flow time series. Lower NMAE values indicate better predictive accuracy, whereas higher $R^2$ values (closer to $1$) indicate that a larger share of the variance in $y_t$ is explained by the model. The NMAE is used to make the error comparable across interconnectors with different flow magnitudes. The metrics are computed jointly across all interconnectors and evaluated time-steps rather than separately per line.

\subsection{Integration into a Reduced PSOM}
\label{sec:SingleCountry}

Lastly, we apply the trained surrogate to \emph{unseen} inputs (demand and VRES time series) to obtain hourly interconnector-flow predictions for the node of interest. We then construct a reduced, single-node LEGO model for the same node. In this reduced PSOM, cross-border exchanges are not optimized endogenously; instead, each interconnector flow is prescribed exogenously and fixed to the surrogate prediction $\hat f_{\ell,t}$ for all $\ell \in \mathcal{L}_{c^*}$ and $t \in \mathcal{T}$. This setup isolates the effect of the imposed flow time series on costs, dispatch, and feasibility (e.g., ENS), because the reduced model uses the same node-specific demand, renewable availability, technology portfolio, and cost parameters as in the full PSOM, while representing the rest of the system only through the fixed interconnector flows. The resulting reduced-model solution is conditionally optimal for the imposed boundary flows, but it is not guaranteed to reproduce the country-level component of the full-system optimum if the predicted flows deviate from the full-model reference.

\added{The quantities evaluated here reflect the intended application of the framework: energy-system planning and scenario analysis, including generation by technology, storage operation, system cost, annual trade, and ENS. The framework is not intended to determine real-time operating points, grid-security margins, or detailed transmission- or distribution-network feasibility, for which the corresponding physics-based network formulation remains mandatory.}

We solve the single-node LEGO model and we record the objective function value $f_{\text{obj}}^{c^*}$, the annual electricity imports and exports, the generation mix by technology with a particular focus on the dispatchable technologies available at the node of interest, and the ENS. We define the relative error in system cost with respect to the reference as
\begin{equation}
\Delta f_{\text{obj}}[\%] = 100 \cdot
\frac{f_{\text{obj}}^{c^*} - f_{\text{obj}}^{c^{*,\mathrm{TRUE}}}}
     {f_{\text{obj}}^{c^{*,\mathrm{TRUE}}}}.
\end{equation}
and we define analogous relative deviations in total annual imports and exports.

\section{Case Study Description}
\label{sec:CaseStudy}

\begin{figure*}
    \centering
    \begin{subfigure}{0.49\textwidth}
        \centering
        \textbf{Full European PSOM}\\[4pt]
        \includegraphics[width=\linewidth]{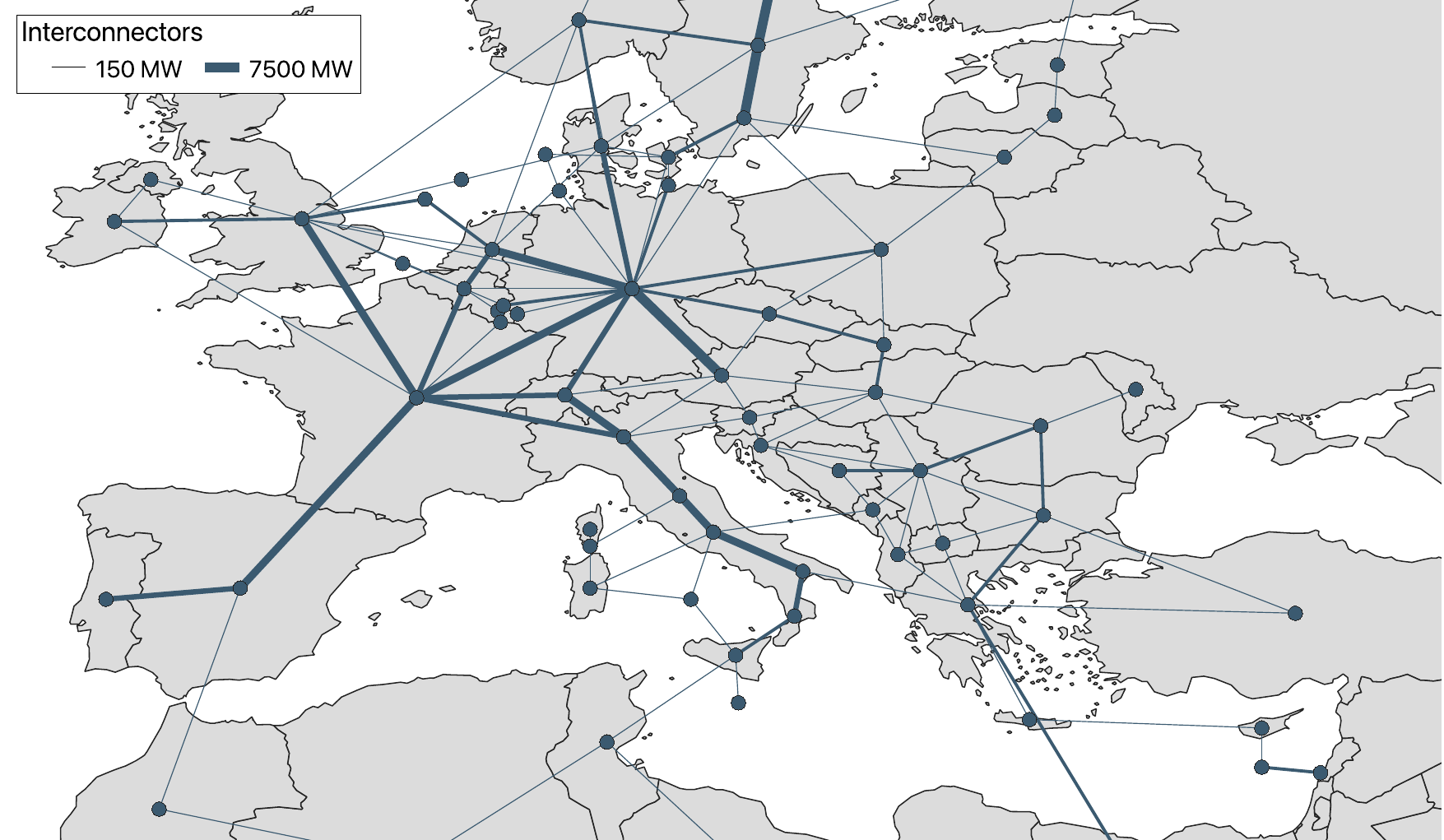}
    \end{subfigure}
    \hfill
    \begin{subfigure}{0.49\textwidth}
        \centering
        \textbf{Single Country PSOM (Austria)}\\[4pt]
        \includegraphics[width=\linewidth]{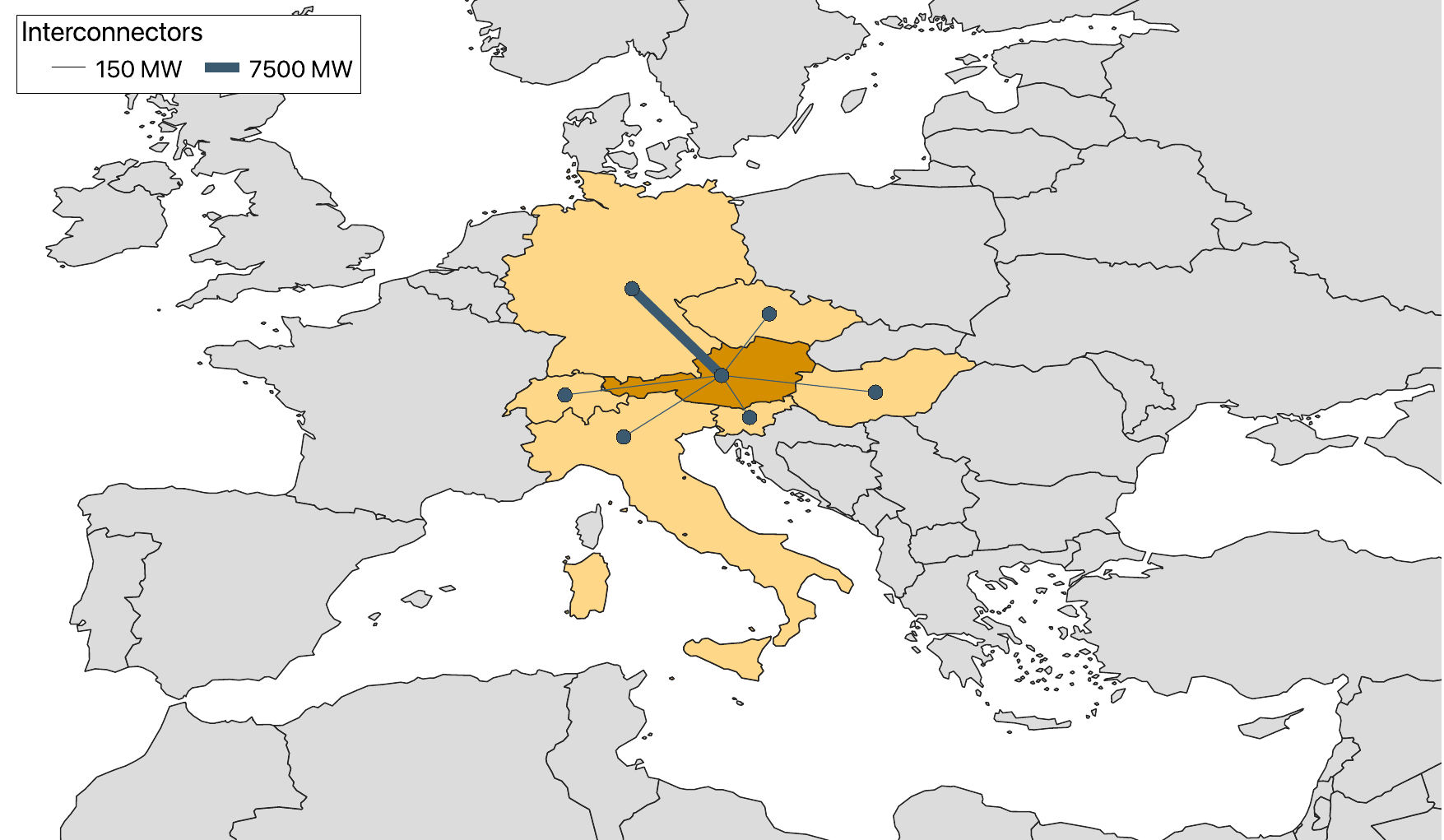}
    \end{subfigure}
    \\[6pt]
    \begin{subfigure}{0.49\textwidth}
        \centering
        \textbf{Single Country PSOM (Germany)}\\[4pt]
        \includegraphics[width=\linewidth]{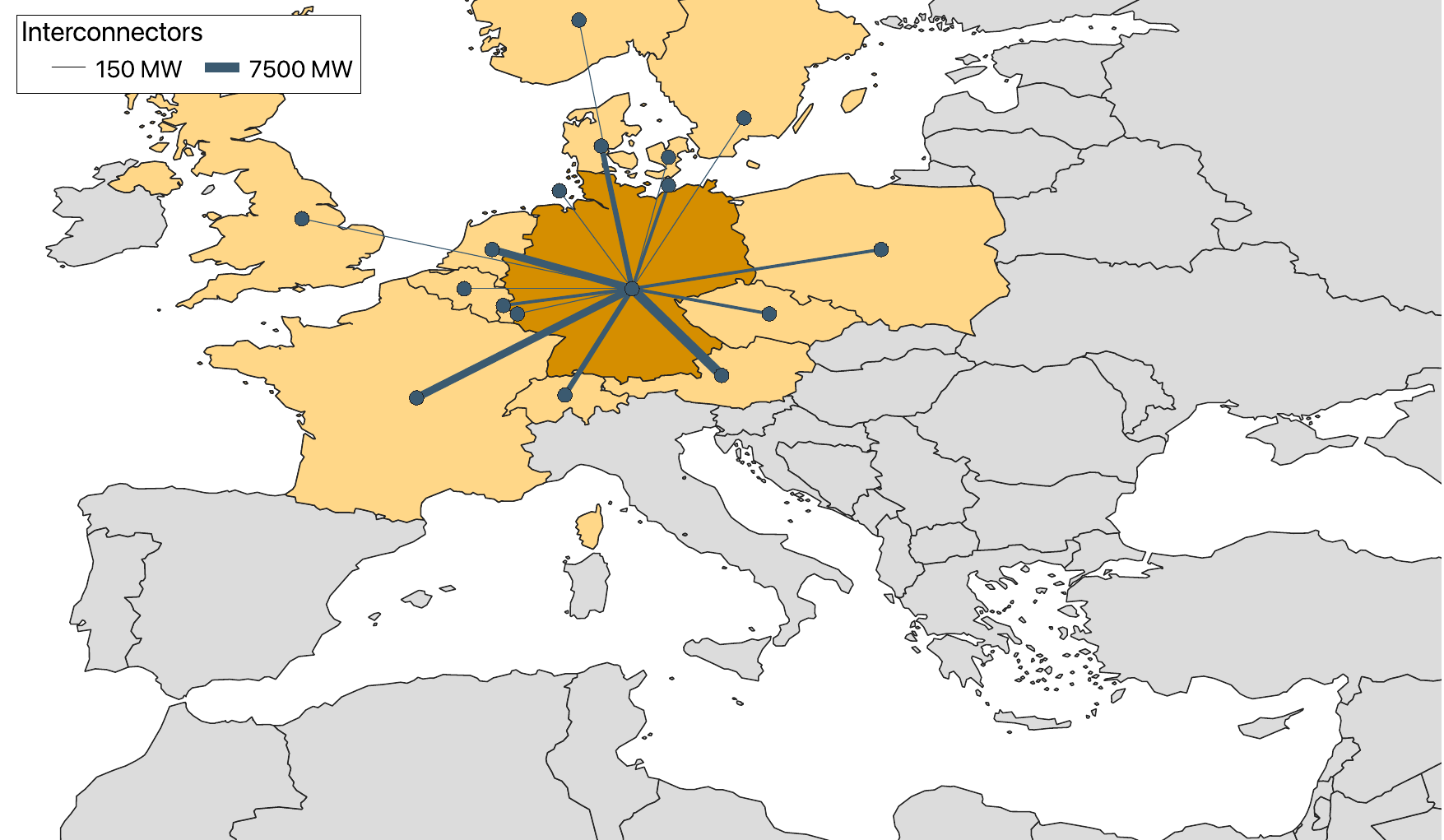}
    \end{subfigure}
    \hfill
    \begin{subfigure}{0.49\textwidth}
        \centering
        \textbf{Single Country PSOM (Spain)}\\[4pt]
        \includegraphics[width=\linewidth]{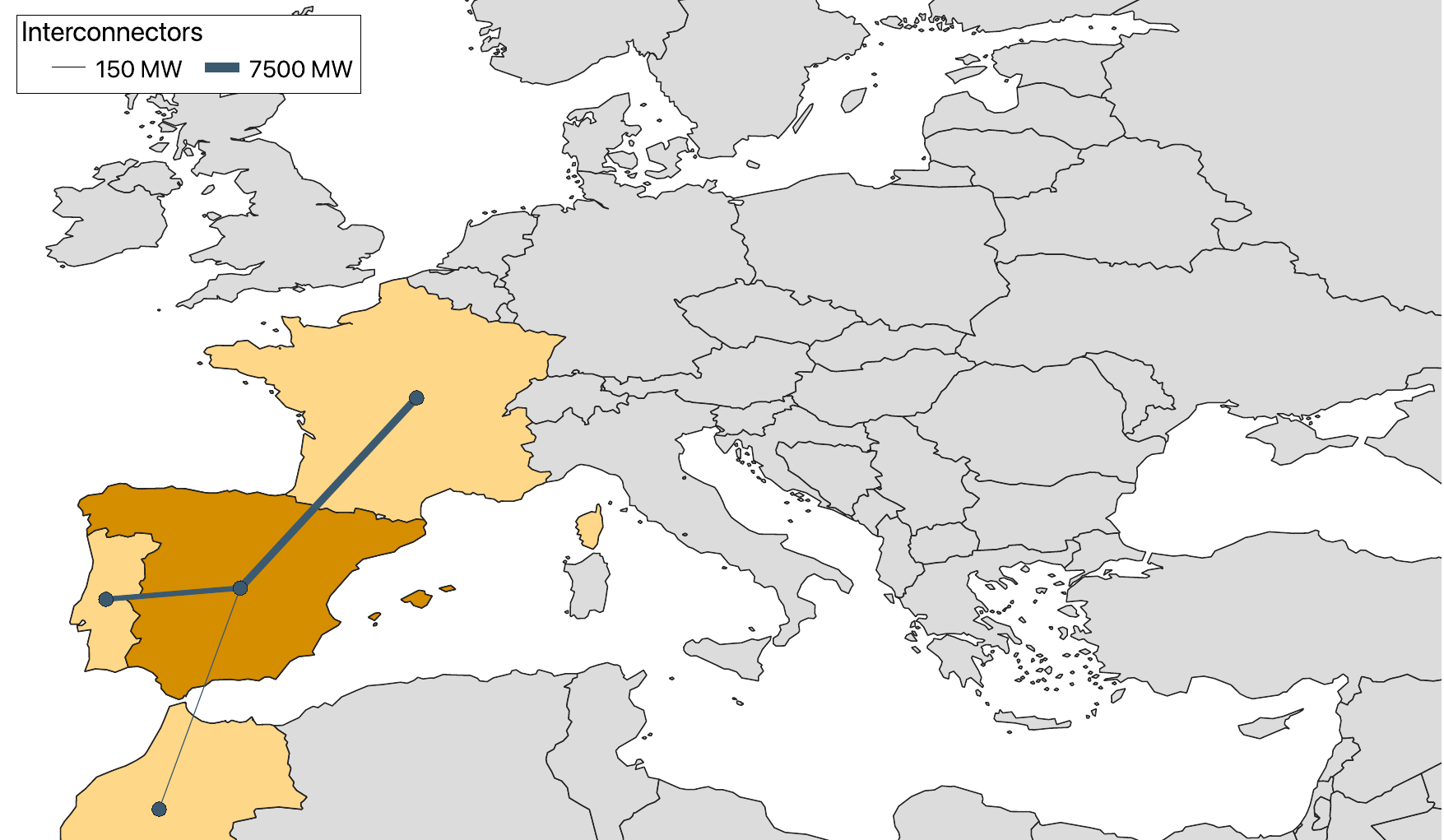}
    \end{subfigure}
    \caption{Spatial scope of the optimization models. Top left: Full European PSOM, representing all modeled countries and cross-border interconnectors. Top right and bottom row: Reduced single-country PSOMs for Austria (top right), Germany (bottom left), and Spain (bottom right). In each reduced PSOM, only the country of interest is optimized explicitly; directly interconnected neighboring countries are shown for geographical context and to indicate the interfaces on which interconnector flows are prescribed exogenously as fixed boundary conditions. Line thickness indicates interconnector capacity (MW). The dark yellow shaded area marks the country of interest, while the light yellow shaded areas indicate its directly interconnected neighboring countries.}
    \label{fig:psom}
\end{figure*}

For the case study, we use a simplified European network using a DC-OPF approach, representing each bidding zone as a single node with hourly resolution in the PSOM (see Figure~\ref{fig:psom}, top left). All techno-economic input data are taken from the ENTSO-E TYNDP 2024 \emph{National Trends} scenario for the year 2030 \citep{entso-e_tyndp_2025}. In most cases, a bidding zone coincides with a single country. However, some countries are subdivided into multiple bidding zones (e.g., Italy, Norway, and Sweden). For simplicity, we refer to bidding zones as \emph{countries} in the remainder of the paper, and thus use $c$ interchangeably as the index for the country/node. For every country~$c$ and hour~$t$, the TYNDP dataset includes hourly electricity demand~$D_{c,t}$ as well as generation time series $G_{c,t}^{k}$ for each non-dispatchable technologies~$k$, such as onshore wind, offshore wind, PV, CSP, run-of-river and other RES. In addition, we incorporate dispatchable generation technologies, including hydro (pumped\textendash) storage units (with inflow profiles), nuclear power plants, gas\textendash fired units, coal\textendash~and oil\textendash based generation, and other non\textendash renewable technologies. Furthermore, we include technical and economic parameters of all technologies such as installed capacity, variable and fixed costs and efficiencies. We also include interconnector capacities between all modeled countries. Since the TYNDP dataset does not provide transmission line reactances required for DC-OPF modeling, we construct a single-node-per-country network representation using the PyPSA data and the PyPSA spatial-resolution clustering method and match the resulting lines with the TYNDP interconnectors \citep{brown_pypsa-eur_2026}.

We select three countries for the case studies: Austria (AT), Germany (DE), and Spain (ES). These countries are chosen to represent distinct structural positions within the European power system. Austria is located at the geographical center of continental Europe and exhibits high levels of cross-border electricity exchange, making it strongly influenced by trans-European power flows. Germany represents one of the largest electricity systems in Europe, both in terms of installed capacity and total generation, and features a highly diversified technology mix with significant shares of VRES. Spain, in contrast, is comparatively weakly interconnected with the rest of the continental European grid, which limits its participation in cross-border electricity trading and results in more localized system dynamics. Together, these three case studies allow us to evaluate the proposed surrogate model under substantially different generation structures and degrees of cross-border integration. 

\added{Austria also illustrates the practical motivation for the proposed framework. In applied studies, a detailed regional or distribution-level model may be the actual object of interest, while its transmission-level boundary conditions remain influenced by Austria's strong coupling to the surrounding European system. Combining an already detailed regional or distribution-grid representation with repeated full pan-European optimizations for many climate years or sensitivities can substantially increase computation time. In such applications, learned cross-border flows can provide scenario-consistent transmission-level boundary information for the geographically reduced analysis. This use case concerns scenario assessment; it does not replace the detailed network calculations required for grid operation or security assessment.}

Using the specified DC-OPF European PSOM, interconnector flows are generated based on the renewable generation profiles (including PV, wind, run-of-river hydro, and CSP) and demand per node for climate year (CY)~2009. This is selected for this case study because the ENTSO-E TYNDP~2024 Implementation Guidelines identify it as the most representative climate year \citep{entso-e_tyndp_2024}. Therefore, a \num{8760} sample dataset is generated that include \num{249} input variables (generation and demand variables). The output dimension depends on the country of interest and equals the number of its interconnectors: \num{6} targets for Austria, \num{16} for Germany, and \num{3} for Spain (each target representing the signed flow on one interconnector; see Table~\ref{tab:model_configurations}). We split these \num{8760} samples into an \SI{80}{\percent}/\SI{20}{\percent} training/test partition to evaluate predictive performance within CY~2009. In addition, we evaluate predictive accuracy and downstream PSOM impacts on the two out-of-sample CYs used in TYNDP~2024 (1995 and 2008), which allows assessing whether the ranking of flow specifications is robust across different weather realizations.

These out-of-sample tests vary the weather-dependent demand and renewable-generation profiles while keeping the structural 2030 scenario assumptions fixed. The surrogate therefore learns the interconnector flows of the full PSOM under the assumed fuel and carbon prices, installed generation and storage capacities, market formulation, transmission capacities, and network topology.

As shown in Figure~\ref{fig:psom}, we construct a single-node LEGO model for each case study country that uses the same technology portfolio, the same demand time series and the same cost parameters as in the full European LEGO model, but that does not explicitly model neighboring nodes. Instead for every cross-border connections $f_{\ell,t}$ , the interconnection flow is fixed exogenously to the values predicted by the respective machine learning model.

To verify the proposed surrogate modeling approach, we evaluate the single-country PSOM using interconnector flows derived from different sources. As a reference, we define the \emph{True} case, which employs the interconnector flows obtained from the full European LEGO simulation for the CY~1995 \& 2008.  
This reference represents the internally consistent outcome of the pan-European model and serves as the benchmark against which all alternatives are assessed.

For the Austrian case study, we further include historical interconnector flows from 2024 as published by the Austrian transmission system operator (\texttt{APG\_2024}). This allows us to assess whether historical cross-border flow time series are suitable boundary conditions for reduced single-country PSOMs. To reflect common practices in single-country modeling studies, we also consider two scaled variants of the APG data: one scaled to match the total annual electricity demand in the 2030 TYNDP scenario (\texttt{APG\_Scaled\_Demand}), and one scaled such that the annual imports/exports equal that of the \emph{True} reference case (\texttt{APG\_Scaled\_ImpExp}).

\section{Results}
\label{sec:Results}

This section presents the results in four parts. First, Section~\ref{sec:results_config} summarizes the evaluated surrogate variants, their effective input dimensionality, and the tuned hyperparameters used in the case study. Second, Section~\ref{sec:results_2009} evaluates predictive performance on the climate year~2009 training/test split. Third, Section~\ref{sec:results_1995} assesses out-of-sample predictive accuracy on the unseen climate years~1995 and~2008. Finally, Section~\ref{sec:results_psom} investigates \emph{decision relevance} by embedding the predicted flow time series as fixed interconnector flows in reduced single-country PSOMs for Austria, Germany, and Spain and comparing objective values (total system cost), annual trade balances, dispatch patterns (including storage operation), feasibility indicators such as ENS, and computational runtimes against the corresponding country components of the full European benchmark.

\subsection{Model Configurations and Hyperparameters}
\label{sec:results_config}

Before discussing predictive performance, Table~\ref{tab:model_configurations} summarizes the evaluated model configurations. \texttt{KNN\_FULL}, \texttt{RIDGE\_FULL}, \texttt{RANDOM\_FORESTS\_FULL}, \texttt{EXTRA\_TREES\_FULL}, \texttt{GRADIENT\_BOOSTING\_FULL}, \texttt{XGBOOST\_FULL}, and \texttt{SQU\_FULL} all use the full input matrix of \num{249} variables. For \texttt{SQU\_SELECTED}, the top \num{20} features per interconnector are retained and their union forms the country-specific reduced feature set. For \texttt{SQU\_OPTIMIZED}, the table reports the best-performing neural-network hyperparameters found by the tuning procedure.

\begin{table}[t]
\caption{Summary of ML model configurations used in the case study. Rows with a white background denote the baseline models, whereas grey-shaded rows denote the SQU variants.}
\label{tab:model_configurations}
\resizebox{\columnwidth}{!}{%
\begin{tabular}{@{}llllll@{}}
\toprule
\textbf{Ctry.} & \textbf{Variant} & \textbf{\#Inp.} & \textbf{\#Outp.} & \textbf{Key hyperparameters} & \textbf{Notes} \\
\midrule
AT & \texttt{KNN\_FULL} & \num{249} & \num{6} & $k=\num{20}$, weights=\texttt{uniform}, alg.=\texttt{auto} & distance-based baseline \\
AT & \texttt{RIDGE\_FULL} & \num{249} & \num{6} & $\alpha=\num{1}$ & linear baseline \\
AT & \texttt{RANDOM\_FOREST\_FULL} & \num{249} & \num{6} & \num{400} trees, unrestricted depth & bagged ensemble \\
AT & \texttt{EXTRA\_TREES\_FULL} & \num{249} & \num{6} & \num{400} trees, unrestricted depth & randomized ensemble \\
AT & \texttt{GRADIENT\_BOOSTING\_FULL} & \num{249} & \num{6} & \num{300} iter., lr=\num{0.05} & histogram gradient boosting \\
AT & \texttt{XGBOOST\_FULL} & \num{249} & \num{6} & \num{500} trees, depth=\num{6}, lr=\num{0.04} & subsample/columns=\num{0.85} \\
\rowcolor{squrow}
AT & \texttt{SQU\_FULL} & \num{249} & \num{6} & epochs=\num{100}, batch=\num{32} & MSE loss \\
\rowcolor{squrow}
AT & \texttt{SQU\_SELECTED} & \num{35} & \num{6} & epochs=\num{100}, batch=\num{32} & top \num{20}/output \\
\rowcolor{squrow}
AT & \texttt{SQU\_OPTIMIZED} & \num{35} & \num{6} & epochs=\num{142}, batch=\num{8} & tuned \\
\rowcolor{squrow}
AT & \texttt{SQU\_CUSTOM\_LOSS} & \num{35} & \num{6} & epochs=\num{142}, batch=\num{8}, +custom loss & feasibility-aware \\
\midrule
DE & \texttt{KNN\_FULL} & \num{249} & \num{16} & $k=\num{20}$, weights=\texttt{uniform}, alg.=\texttt{auto} & distance-based baseline \\
DE & \texttt{RIDGE\_FULL} & \num{249} & \num{16} & $\alpha=\num{1}$ & linear baseline \\
DE & \texttt{RANDOM\_FOREST\_FULL} & \num{249} & \num{16} & \num{400} trees, unrestricted depth & bagged ensemble \\
DE & \texttt{EXTRA\_TREES\_FULL} & \num{249} & \num{16} & \num{400} trees, unrestricted depth & randomized ensemble \\
DE & \texttt{GRADIENT\_BOOSTING\_FULL} & \num{249} & \num{16} & \num{300} iter., lr=\num{0.05} & histogram gradient boosting \\
DE & \texttt{XGBOOST\_FULL} & \num{249} & \num{16} & \\num{500} trees, depth=\num{6}, lr=\num{0.04} & subsample/columns=\num{0.85} \\
\rowcolor{squrow}
DE & \texttt{SQU\_FULL} & \num{249} & \num{16} & epochs=\num{100}, batch=\num{32} & MSE loss \\
\rowcolor{squrow}
DE & \texttt{SQU\_SELECTED} & \num{43} & \num{16} & epochs=\num{100}, batch=\num{32} & top \num{20}/output \\
\rowcolor{squrow}
DE & \texttt{SQU\_OPTIMIZED} & \num{43} & \num{16} & epochs=\num{142}, batch=\num{8} & tuned \\
\rowcolor{squrow}
DE & \texttt{SQU\_CUSTOM\_LOSS} & \num{43} & \num{16} & epochs=\num{142}, batch=\num{8}, +custom loss & feasibility-aware \\
\midrule
ES & \texttt{KNN\_FULL} & \num{249} & \num{3} & $k=\num{20}$, weights=\texttt{uniform}, alg.=\texttt{auto} & distance-based baseline \\
ES & \texttt{RIDGE\_FULL} & \num{249} & \num{3} & $\alpha=\num{1}$ & linear baseline \\
ES & \texttt{RANDOM\_FOREST\_FULL} & \num{249} & \num{3} & \num{400} trees, unrestricted depth & bagged ensemble \\
ES & \texttt{EXTRA\_TREES\_FULL} & \num{249} & \num{3} & \num{400} trees, unrestricted depth & randomized ensemble \\
ES & \texttt{GRADIENT\_BOOSTING\_FULL} & \num{249} & \num{3} & \num{300} iter., lr=\num{0.05} & histogram gradient boosting \\
ES & \texttt{XGBOOST\_FULL} & \num{249} & \num{3} & \num{500} trees, depth=\num{6}, lr=\num{0.04} & subsample/columns=\num{0.85} \\
\rowcolor{squrow}
ES & \texttt{SQU\_FULL} & \num{249} & \num{3} & epochs=\num{100}, batch=\num{32} & MSE loss \\
\rowcolor{squrow}
ES & \texttt{SQU\_SELECTED} & \num{25} & \num{3} & epochs=\num{100}, batch=\num{32} & top \num{20}/output \\
\rowcolor{squrow}
ES & \texttt{SQU\_OPTIMIZED} & \num{25} & \num{3} & epochs=\num{142}, batch=\num{8} & tuned \\
\rowcolor{squrow}
ES & \texttt{SQU\_CUSTOM\_LOSS} & \num{25} & \num{3} & epochs=\num{142}, batch=\num{8}, +custom loss & feasibility-aware \\
\bottomrule
\end{tabular}
}
\end{table}

\subsection{Training and Testing Results of Interconnector Flows for CY 2009}
\label{sec:results_2009}

Table~\ref{tab:metrics_train_test_all} reports the overall NMAE and $R^2$ values for the 2009 training and test sets. 

The baselines reveal that nonlinear tree ensembles are highly competitive on the within-year test split. For Austria (AT) and Germany (DE), \texttt{EXTRA\_TREES\_FULL} and \texttt{XGBOOST\_FULL} achieve the lowest test NMAE values (\numrange{0.21}{0.22}) and test $R^2$ values between \num{0.87} and \num{0.90}. \texttt{GRADIENT\_BOOSTING\_FULL} is similarly strong, with test NMAE of \num{0.23} for Austria and \num{0.22} for Germany. Random forests reach test NMAE of \num{0.25}--\num{0.26}, whereas ridge regression remains close to \texttt{KNN\_FULL}, indicating that a purely linear mapping is insufficient.

For Spain, the strongest tree-based baselines reach test NMAE values of approximately \numrange{0.32}{0.35} and test $R^2$ values of \numrange{0.77}{0.79}. These results are comparable to the SQU variants, whose test NMAE ranges from \num{0.31} to \num{0.33}. The perfect training fit of \texttt{EXTRA\_TREES\_FULL} ($R^2=\num{1.00}$) and the pronounced train--test gap of several tree ensembles nevertheless indicate substantial in-sample memorization, motivating the out-of-sample climate-year evaluation in Section~\ref{sec:results_1995}.

Overall, the SQU variants achieve predictive performance comparable to the strongest tree-based ensembles on the 2009 test split. While \texttt{EXTRA\_TREES\_FULL}, \texttt{XGBOOST\_FULL}, and \texttt{GRADIENT\_BOOSTING\_FULL} obtain the lowest statistical errors for Austria and Germany, the SQU models consistently rank among the best-performing approaches across all three case studies. \texttt{SQU\_CUSTOM\_LOSS} illustrates the trade-off between feasibility-aware training and statistical accuracy. For AT and ES its predictive performance remains close to the other SQU variants, but for DE it is notably worse (test $R^2=\num{0.63}$, test NMAE \num{0.31}), indicating that the additional penalty can distort the learned mapping in systems where the reference data from the full European PSOM itself contains small ENS for Germany that the custom loss tends to suppress. 

Consequently, the within-year results demonstrate that SQU provides a robust approximation of interconnector flows that is competitive with state-of-the-art tree ensembles, whereas the relative merits of the different approaches must ultimately be assessed using unseen climate years and their downstream impact on reduced PSOM results.

\newcommand{\nmaecell}[1]{%
  \begingroup
  \pgfmathtruncatemacro{\class}{(#1 < 0.40) ? 1 : ((#1 > 0.60) ? 3 : 2)}%
  \ifnum\class=1 \cellcolor{good}\fi
  \ifnum\class=2 \cellcolor{mid}\fi
  \ifnum\class=3 \cellcolor{bad}\fi
  \num{#1}%
  \endgroup
}

\newcommand{\rTwocell}[1]{%
  \begingroup
  \pgfmathtruncatemacro{\class}{(#1 >= 0.50) ? 1 : ((#1 >= 0.30) ? 2 : 3)}%
  \ifnum\class=1 \cellcolor{good}\fi
  \ifnum\class=2 \cellcolor{mid}\fi
  \ifnum\class=3 \cellcolor{bad}\fi
  \num{#1}%
  \endgroup
}

\begin{table}[t]
\caption{Overall NMAE and $R^2$ on the 2009 training and test sets for the three case-study countries. Lower NMAE and higher $R^2$ values (closer to 1) indicate better predictive accuracy. Cell background indicates performance (green: good, yellow: intermediate, red: weak).}
\label{tab:metrics_train_test_all}
\resizebox{\columnwidth}{!}{%
\begin{tabular}{@{}llrrrrrr@{}}
\toprule
\multicolumn{2}{c}{\multirow{2}{*}{Model}} & \multicolumn{2}{c}{AT} & \multicolumn{2}{c}{DE} & \multicolumn{2}{c}{ES} \\ \cmidrule(l){3-8}
\multicolumn{2}{c}{} & Train & Test & Train & Test & Train & Test \\ \midrule
\multirow{2}{*}{\texttt{KNN\_FULL}}
& NMAE & \nmaecell{0.37} & \nmaecell{0.40} & \nmaecell{0.38} & \nmaecell{0.41} & \nmaecell{0.53} & \nmaecell{0.55} \\
& Overall $R^2$ & \rTwocell{0.75} & \rTwocell{0.71} & \rTwocell{0.73} & \rTwocell{0.70} & \rTwocell{0.58} & \rTwocell{0.54} \\ \midrule
\multirow{2}{*}\texttt{RIDGE\_FULL}
& NMAE & \nmaecell{0.38} & \nmaecell{0.40} & \nmaecell{0.37} & \nmaecell{0.39} & \nmaecell{0.53} & \nmaecell{0.54} \\
& Overall $R^2$ & \rTwocell{0.72} & \rTwocell{0.70} & \rTwocell{0.77} & \rTwocell{0.75} & \rTwocell{0.56} & \rTwocell{0.54} \\ \midrule
\multirow{2}{*}\texttt{RANDOM\_FOREST\_FULL}
& NMAE & \nmaecell{0.09} & \nmaecell{0.25} & \nmaecell{0.09} & \nmaecell{0.26} & \nmaecell{0.13} & \nmaecell{0.35} \\
& Overall $R^2$ & \rTwocell{0.98} & \rTwocell{0.85} & \rTwocell{0.98} & \rTwocell{0.85} & \rTwocell{0.97} & \rTwocell{0.77} \\ \midrule
\multirow{2}{*}\texttt{EXTRA\_TREES\_FULL}
& NMAE & \nmaecell{0.00} & \nmaecell{0.22} & \nmaecell{0.00} & \nmaecell{0.21} & \nmaecell{0.00} & \nmaecell{0.32} \\
& Overall $R^2$ & \rTwocell{1.00} & \rTwocell{0.88} & \rTwocell{1.00} & \rTwocell{0.89} & \rTwocell{1.00} & \rTwocell{0.79} \\ \midrule
\multirow{2}{*}\texttt{GRADIENT\_BOOSTING\_FULL}
& NMAE & \nmaecell{0.12} & \nmaecell{0.23} & \nmaecell{0.11} & \nmaecell{0.22} & \nmaecell{0.17} & \nmaecell{0.32} \\
& Overall $R^2$ & \rTwocell{0.96} & \rTwocell{0.87} & \rTwocell{0.98} & \rTwocell{0.89} & \rTwocell{0.95} & \rTwocell{0.79} \\ \midrule
\multirow{2}{*}\texttt{XGBOOST\_FULL}
& NMAE & \nmaecell{0.08} & \nmaecell{0.22} & \nmaecell{0.07} & \nmaecell{0.21} & \nmaecell{0.12} & \nmaecell{0.32} \\
& Overall $R^2$ & \rTwocell{0.98} & \rTwocell{0.87} & \rTwocell{0.99} & \rTwocell{0.90} & \rTwocell{0.97} & \rTwocell{0.79} \\ \midrule
\multirow{2}{*}{\texttt{SQU\_FULL}}
& NMAE & \nmaecell{0.16} & \nmaecell{0.24} & \nmaecell{0.18} & \nmaecell{0.24} & \nmaecell{0.19} & \nmaecell{0.33} \\
& Overall $R^2$ & \rTwocell{0.93} & \rTwocell{0.85} & \rTwocell{0.92} & \rTwocell{0.85} & \rTwocell{0.90} & \rTwocell{0.70} \\ \midrule
\multirow{2}{*}{\texttt{SQU\_SELECTED}}
& NMAE & \nmaecell{0.19} & \nmaecell{0.26} & \nmaecell{0.20} & \nmaecell{0.24} & \nmaecell{0.21} & \nmaecell{0.32} \\
& Overall $R^2$ & \rTwocell{0.91} & \rTwocell{0.82} & \rTwocell{0.90} & \rTwocell{0.85} & \rTwocell{0.88} & \rTwocell{0.75} \\ \midrule
\multirow{2}{*}{\texttt{SQU\_OPTIMIZED}}
& NMAE & \nmaecell{0.19} & \nmaecell{0.26} & \nmaecell{0.20} & \nmaecell{0.24} & \nmaecell{0.19} & \nmaecell{0.31} \\
& Overall $R^2$ & \rTwocell{0.91} & \rTwocell{0.82} & \rTwocell{0.91} & \rTwocell{0.86} & \rTwocell{0.90} & \rTwocell{0.74} \\ \midrule
\multirow{2}{*}{\texttt{SQU\_CUSTOM\_LOSS}}
& NMAE & \nmaecell{0.21} & \nmaecell{0.29} & \nmaecell{0.30} & \nmaecell{0.31} & \nmaecell{0.21} & \nmaecell{0.32} \\
& Overall $R^2$ & \rTwocell{0.91} & \rTwocell{0.82} & \rTwocell{0.64} & \rTwocell{0.63} & \rTwocell{0.88} & \rTwocell{0.71} \\
\bottomrule
\end{tabular}
}
\end{table}

\subsection{Prediction of Cross-Border Interconnector Flows for the CY 1995 \& 2008}
\label{sec:results_1995}

\begin{figure*}
    \centering
    \includegraphics[width=1\linewidth]{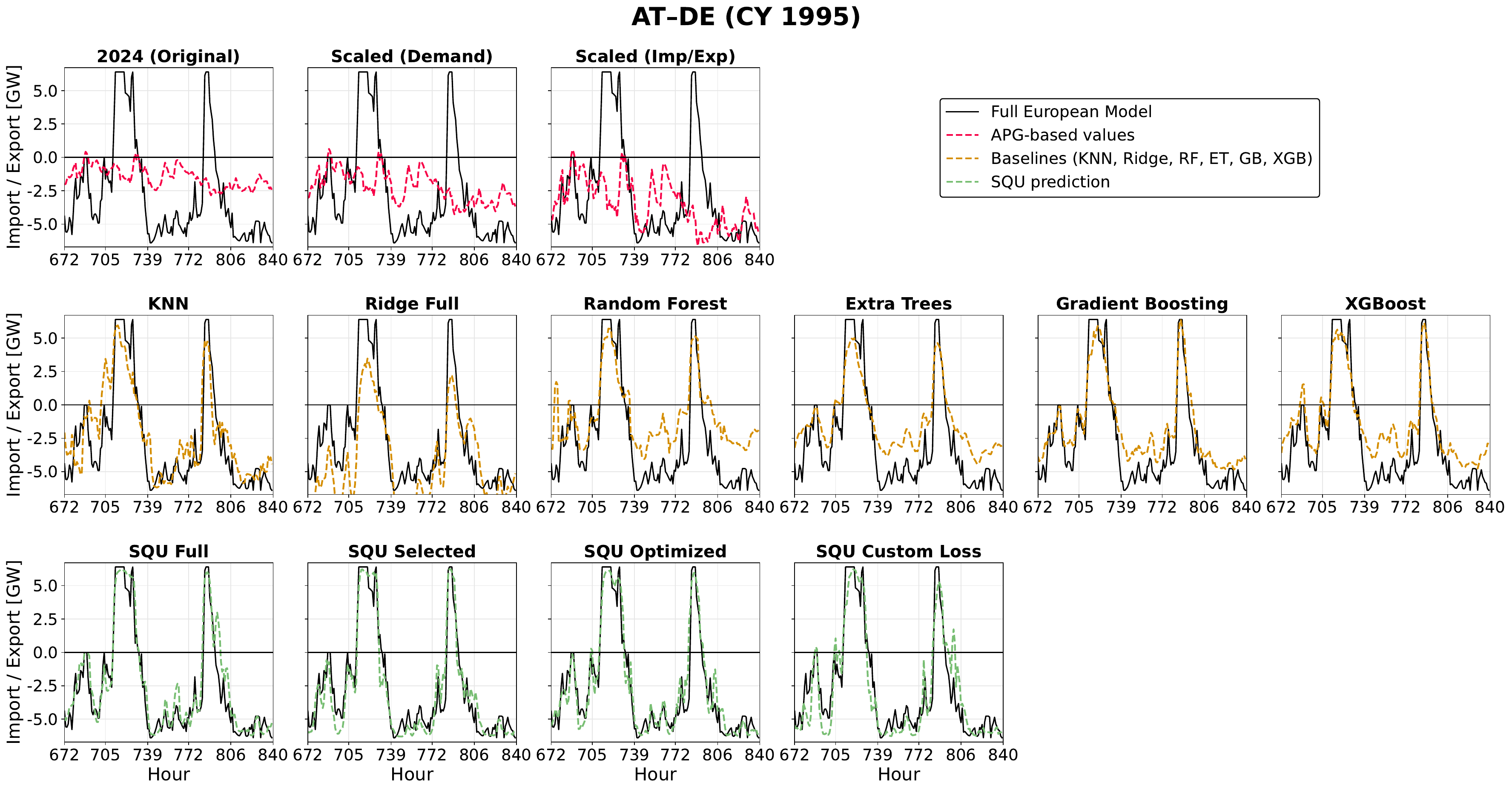}
    \caption{Comparison of hourly cross-border interconnector flows between Austria and Germany for week~5 of climate year~1995. Positive values indicate imports into Austria and negative values exports. The black line is the full European LEGO PSOM reference. Red lines show the APG~2024 historical profile and two scaled variants. Orange lines show the full-feature tabular baselines: KNN, ridge regression, random forest, Extra Trees, gradient boosting, and XGBoost. Green lines show the SQU variants using full, selected, optimized, and custom-loss configurations.}
    \label{fig:AT–DE_week_5}
\end{figure*}

Figure~\ref{fig:AT–DE_week_5} shows hourly cross-border flows between Austria and Germany for one representative week. The \textbf{black} curve is the \texttt{Full European Model} (full LEGO PSOM) reference for the out-of-sample climate year~1995, i.e., the benchmark flow trajectory produced endogenously by the pan-European DC-OPF.

To illustrate the limitations of common ``fixed-flow'' practice, the \textbf{\textcolor[HTML]{F70147}{red}} curves use historical exchanges published by the Austrian TSO APG for the calendar year~2024. These APG-based series are \emph{not} climate-year-consistent with the 1995 weather realization; they are included purely as benchmark boundary conditions that practitioners often reuse when running reduced (single-country) models. Besides the original 2024 series, we also show two scaled variants: one scaled to match the 2030 scenario demand level (``Scaled (Demand)'') and one scaled to match annual import/export totals (``Scaled (Imp/Exp)'').

The remaining panels show the ML surrogates applied to the same climate year~1995 inputs. In the \textbf{\textcolor[HTML]{D58E00}{orange}} row (Baselines) and the \textbf{\textcolor[HTML]{78BE73}{green}} row (SQU), the dashed curves are the predicted flows for the different model variants, each compared against the same \textbf{black} full-model reference. This visualization highlights that matching annual totals via scaling does not recover the correct hourly structure, whereas the ML surrogates better reproduce the timing and magnitude of the 1995 flow pattern. The tree ensembles reproduce the principal flow peaks well, although noticeable deviations remain during low-flow periods and around sign changes, which are particularly relevant for the downstream reduced PSOM.

Across the two out-of-sample CYs (1995 and 2008), Table~\ref{tab:metrics_1995_2008_all} confirms these qualitative observations. For Austria, the APG baselines perform very poorly in both years, with NMAE around \numrange{1.09}{1.32} and strongly negative $R^2$ values (e.g., $R^2=\num{-1.21}$ in 1995 and $R^2=\num{-1.44}$ in 2008 for \texttt{APG\_2024}), indicating performance worse than a trivial mean predictor. The benchmark shows a clear hierarchy among the baseline models. \texttt{RIDGE\_FULL} and \texttt{KNN\_FULL} generalize least robustly, whereas the tree ensembles substantially reduce the error. \texttt{GRADIENT\_BOOSTING\_FULL} is the strongest conventional baseline overall: its NMAE ranges from \num{0.33} to \num{0.44} and its $R^2$ from \num{0.66} to \num{0.75} across the six country--year combinations. \texttt{XGBOOST\_FULL} follows closely, with NMAE of \numrange{0.34}{0.46} and $R^2$ of \numrange{0.65}{0.74}.

While no single model class consistently achieves the highest predictive accuracy across all countries, the SQU variants deliver the most consistently strong performance across all three test regions. They remain competitive with the best-performing tree ensembles in Austria and Germany while achieving the highest accuracy in Spain.

\renewcommand{\nmaecell}[1]{%
  \begingroup
  \pgfmathtruncatemacro{\class}{(#1 < 0.40) ? 1 : ((#1 > 0.60) ? 3 : 2)}%
  \ifnum\class=1 \cellcolor{good}\fi
  \ifnum\class=2 \cellcolor{mid}\fi
  \ifnum\class=3 \cellcolor{bad}\fi
  \num{#1}%
  \endgroup
}

\renewcommand{\rTwocell}[1]{%
  \begingroup
  \pgfmathtruncatemacro{\class}{(#1 >= 0.50) ? 1 : ((#1 >= 0.30) ? 2 : 3)}%
  \ifnum\class=1 \cellcolor{good}\fi
  \ifnum\class=2 \cellcolor{mid}\fi
  \ifnum\class=3 \cellcolor{bad}\fi
  \num{#1}%
  \endgroup
}

\begin{table}[t]
\caption{Overall NMAE and $R^2$ on the out-of-sample climate years 1995 and 2008 for the three case-study countries, including the five additional full-feature baselines. Lower NMAE and higher $R^2$ (closer to 1) indicate better performance. Cell background indicates performance (green: good, yellow: intermediate, red: weak).}
\label{tab:metrics_1995_2008_all}
\resizebox{\columnwidth}{!}{%
\begin{tabular}{@{}llrrrrrr@{}}
\toprule
\multirow{2}{*}{Model} & \multirow{2}{*}{Score} &
\multicolumn{3}{c}{\textbf{1995}} &
\multicolumn{3}{c}{\textbf{2008}} \\
\cmidrule(lr){3-5}\cmidrule(lr){6-8}
& & AT & DE & ES & AT & DE & ES \\
\midrule
\multirow{2}{*}{\texttt{APG\_2024}}
& NMAE & \nmaecell{1.09} & n.a. & n.a. & \nmaecell{1.10} & n.a. & n.a. \\
& Overall $R^2$ & \rTwocell{-1.21} & n.a. & n.a. & \rTwocell{-1.44} & n.a. & n.a. \\ \midrule
\multirow{2}{*}{\texttt{APG\_Scaled\_Demand}}
& NMAE & \nmaecell{1.24} & n.a. & n.a. & \nmaecell{1.26} & n.a. & n.a. \\
& Overall $R^2$ & \rTwocell{-2.57} & n.a. & n.a. & \rTwocell{-2.96} & n.a. & n.a. \\ \midrule
\multirow{2}{*}{\texttt{APG\_Scaled\_ImpExp}}
& NMAE & \nmaecell{1.30} & n.a. & n.a. & \nmaecell{1.32} & n.a. & n.a. \\
& Overall $R^2$ & \rTwocell{-3.79} & n.a. & n.a. & \rTwocell{-4.51} & n.a. & n.a. \\ \midrule
\multirow{2}{*}{\texttt{KNN\_FULL}}
& NMAE & \nmaecell{0.54} & \nmaecell{0.58} & \nmaecell{0.71} & \nmaecell{0.51} & \nmaecell{0.56} & \nmaecell{0.72} \\
& Overall $R^2$ & \rTwocell{0.47} & \rTwocell{0.45} & \rTwocell{0.29} & \rTwocell{0.47} & \rTwocell{0.47} & \rTwocell{0.25} \\ \midrule
\multirow{2}{*}\texttt{RIDGE\_FULL}
& NMAE & \nmaecell{0.54} & \nmaecell{0.52} & \nmaecell{0.67} & \nmaecell{0.52} & \nmaecell{0.51} & \nmaecell{0.67} \\
& Overall $R^2$ & \rTwocell{0.39} & \rTwocell{0.50} & \rTwocell{0.19} & \rTwocell{0.42} & \rTwocell{0.53} & \rTwocell{0.28} \\ \midrule
\multirow{2}{*}\texttt{RANDOM\_FOREST\_FULL}
& NMAE & \nmaecell{0.47} & \nmaecell{0.42} & \nmaecell{0.46} & \nmaecell{0.44} & \nmaecell{0.41} & \nmaecell{0.48} \\
& Overall $R^2$ & \rTwocell{0.57} & \rTwocell{0.62} & \rTwocell{0.68} & \rTwocell{0.57} & \rTwocell{0.64} & \rTwocell{0.64} \\ \midrule
\multirow{2}{*}\texttt{EXTRA\_TREES\_FULL}
& NMAE & \nmaecell{0.45} & \nmaecell{0.42} & \nmaecell{0.47} & \nmaecell{0.43} & \nmaecell{0.41} & \nmaecell{0.49} \\
& Overall $R^2$ & \rTwocell{0.60} & \rTwocell{0.69} & \rTwocell{0.68} & \rTwocell{0.61} & \rTwocell{0.65} & \rTwocell{0.63} \\ \midrule
\multirow{2}{*}\texttt{GRADIENT\_BOOSTING\_FULL}
& NMAE & \nmaecell{0.38} & \nmaecell{0.34} & \nmaecell{0.41} & \nmaecell{0.36} & \nmaecell{0.33} & \nmaecell{0.44} \\
& Overall $R^2$ & \rTwocell{0.66} & \rTwocell{0.73} & \rTwocell{0.72} & \rTwocell{0.67} & \rTwocell{0.75} & \rTwocell{0.67} \\ \midrule
\multirow{2}{*}\texttt{XGBOOST\_FULL}
& NMAE & \nmaecell{0.39} & \nmaecell{0.35} & \nmaecell{0.43} & \nmaecell{0.37} & \nmaecell{0.34} & \nmaecell{0.46} \\
& Overall $R^2$ & \rTwocell{0.65} & \rTwocell{0.73} & \rTwocell{0.70} & \rTwocell{0.66} & \rTwocell{0.74} & \rTwocell{0.66} \\ \midrule
\multirow{2}{*}{\texttt{SQU\_FULL}}
& NMAE & \nmaecell{0.39} & \nmaecell{0.43} & \nmaecell{0.46} & \nmaecell{0.38} & \nmaecell{0.42} & \nmaecell{0.49} \\
& Overall $R^2$ & \rTwocell{0.59} & \rTwocell{0.61} & \rTwocell{0.54} & \rTwocell{0.59} & \rTwocell{0.63} & \rTwocell{0.48} \\ \midrule
\multirow{2}{*}{\texttt{SQU\_SELECTED}}
& NMAE & \nmaecell{0.38} & \nmaecell{0.41} & \nmaecell{0.37} & \nmaecell{0.37} & \nmaecell{0.38} & \nmaecell{0.39} \\
& Overall $R^2$ & \rTwocell{0.57} & \rTwocell{0.65} & \rTwocell{0.68} & \rTwocell{0.57} & \rTwocell{0.67} & \rTwocell{0.64} \\ \midrule
\multirow{2}{*}{\texttt{SQU\_OPTIMIZED}}
& NMAE & \nmaecell{0.38} & \nmaecell{0.41} & \nmaecell{0.36} & \nmaecell{0.37} & \nmaecell{0.37} & \nmaecell{0.41} \\
& Overall $R^2$ & \rTwocell{0.59} & \rTwocell{0.65} & \rTwocell{0.69} & \rTwocell{0.57} & \rTwocell{0.70} & \rTwocell{0.62} \\ \midrule
\multirow{2}{*}{\texttt{SQU\_CUSTOM\_LOSS}}
& NMAE & \nmaecell{0.42} & \nmaecell{0.44} & \nmaecell{0.36} & \nmaecell{0.40} & \nmaecell{0.41} & \nmaecell{0.40} \\
& Overall $R^2$ & \rTwocell{0.55} & \rTwocell{0.45} & \rTwocell{0.69} & \rTwocell{0.57} & \rTwocell{0.54} & \rTwocell{0.62} \\
\bottomrule
\end{tabular}
}
\end{table}

\subsubsection{Annual Import and Export Balances}
\label{sec:results_psom_trade}

\begin{figure*}
    \centering
    \includegraphics[width=1\linewidth]{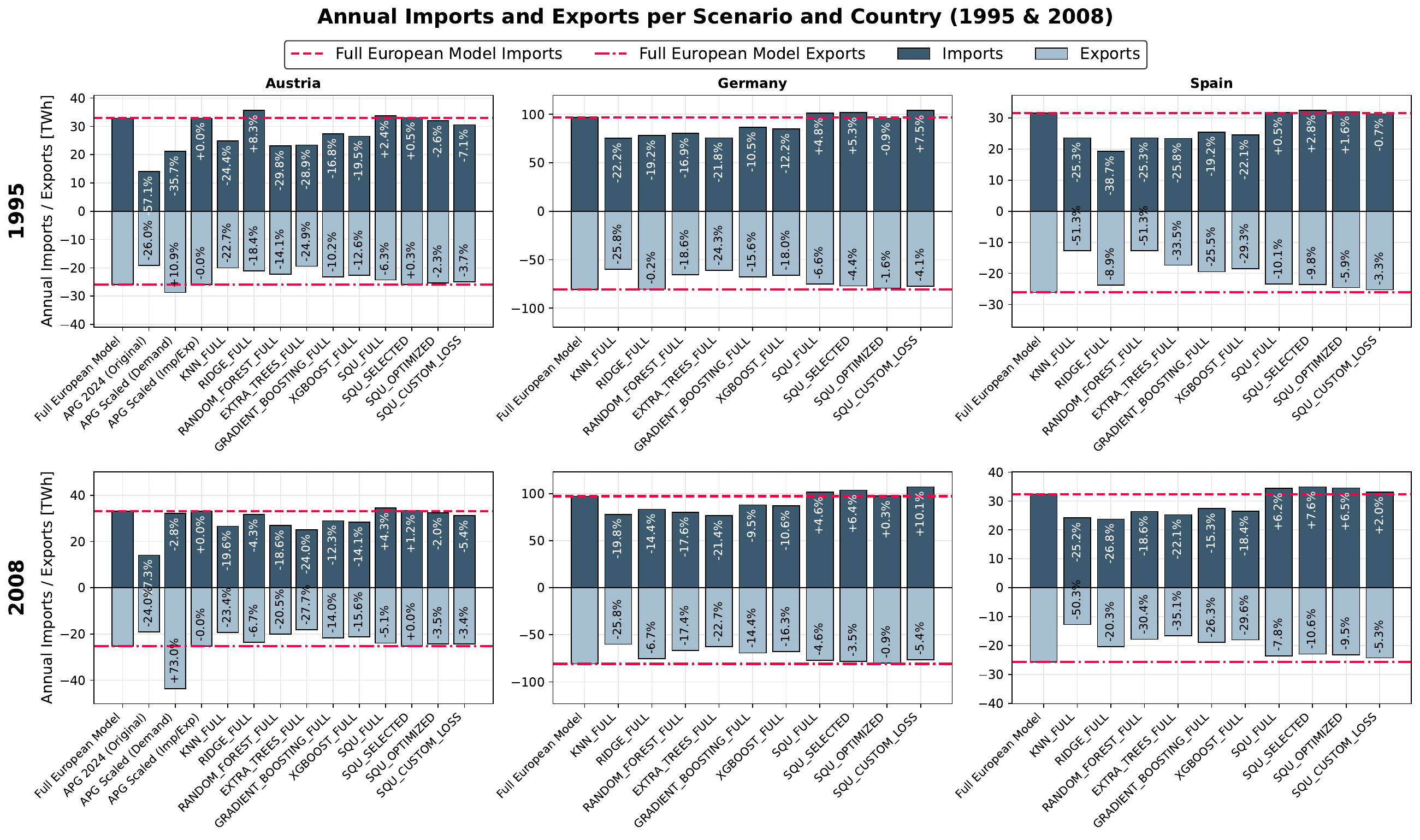}
    \caption{Annual imports and exports in the single-country PSOM for Austria (AT), Germany (DE), and Spain (ES) under alternative cross-border flow specifications for the two out-of-sample climate years 1995 and 2008. Imports are shown as positive values and exports as negative values. Dashed lines indicate the \texttt{Full European PSOM Model} reference.}
    \label{fig:psom_trade_dev}
\end{figure*}

Annual import and export volumes provide a complementary consistency check on the implied cross-border interaction when cross-border exchanges are prescribed exogenously. Figure~\ref{fig:psom_trade_dev} compares annual imports (positive bars) and exports (negative bars) of the country of interest in the reduced single-node PSOM against the corresponding country-level imports and exports from the \texttt{Full European PSOM Model} benchmark (dashed lines) for the two out-of-sample climate years 1995 and 2008 and the three case-study countries.

The APG-based benchmark series for Austria again highlight the limitations of reusing historical exchange patterns under structurally different system conditions. For both out-of-sample years, the unscaled APG~2024 profile produces very large deviations from the benchmark (e.g., imports are underestimated by roughly \SI{-57}{\percent} in both 1995 and 2008). The two scaled variants illustrate different limitations: \texttt{APG\_Scaled\_Demand} adjusts the historical profile to the 2030 demand level but does not necessarily reproduce annual trade volumes, while \texttt{APG\_Scaled\_ImpExp} matches annual import/export totals by construction. As shown earlier in Figure~\ref{fig:AT–DE_week_5}, even when annual trade totals are forced to coincide, the hourly trajectory can remain inconsistent with the benchmark, and it is precisely this temporal distribution that drives dispatch, storage cycling, and cost formation.

Across both out-of-sample years, the ML-based surrogates preserve annual trade volumes more consistently than the historical APG-based profiles. \texttt{RIDGE\_FULL} and the tree ensembles sometimes reproduce annual totals closely, but no single conventional baseline dominates across all countries, years, imports, and exports. The SQU variants remain comparatively close to the full-model reference in most panels. For Austria, the SQU variants keep import and export deviations within single-digit percentages. In 1995, import deviations range from approximately \SI[retain-explicit-plus=true]{+0.5}{\percent} to \SI{-7.1}{\percent}, while export deviations range from approximately \SI[retain-explicit-plus=true]{+0.3}{\percent} to \SI{-3.7}{\percent}. A similar pattern holds in 2008, where the SQU variants remain close to the benchmark, whereas the full-feature baseline models generally exhibit larger negative import deviations.

For Germany and Spain, the same two-year comparison shows that SQU variants typically reproduce annual imports and exports more closely than the conventional full-feature baselines. In Germany, the baseline models imply sizeable import and export shortfalls in both years, whereas SQU variants reduce these deviations to small single-digit values in most cases (with \texttt{SQU\_CUSTOM\_LOSS} standing out as an exception due to its systematically higher implied imports). In Spain, the baseline models again exhibit larger import underestimation, while SQU variants cluster nearer to the benchmark for both imports and exports across 1995 and 2008.

Overall, the two out-of-sample years reinforce the central point: matching annual totals ex post (as in scaling approaches) is not sufficient for economic consistency, whereas the ML-based surrogates, particularly the SQU family, better preserve the cross-border interaction implied by the full European benchmark in both magnitude and (as shown by the weekly comparison) temporal structure.

\subsection{Impact on PSOM Results}
\label{sec:results_psom}

The statistical evaluation in the previous sections shows that ML-based surrogates can reproduce interconnector-flow patterns substantially better than reusing historical (and scaled) profiles, which is valuable in its own right for generating scenario-consistent flow time series. Beyond prediction accuracy, however, the same surrogates can also serve as a practical modeling device in reduced spatial PSOMs: by prescribing ML-generated flows as boundary conditions, we aim to retain the key influence of cross-border exchanges while avoiding repeated full European solves. 

We therefore assess the surrogate models based on their impact on key optimization outputs. To this end, we run single-country LEGO models for Austria, Germany, and Spain under alternative cross-border flow specifications. The comparison includes three classes of flow representations: (i) the \texttt{True} reference derived from the full European LEGO model for climate years 1995 and 2008, (ii) ML-generated interconnector flows from KNN, ridge regression, random forests, Extra Trees, gradient boosting, XGBoost, and the SQU variants for climate years 1995 and 2008, and (iii) for Austria only, APG-based benchmark series constructed from observed 2024 flows (original and scaled variants). For each country, we assess deviations in total system cost (objective function) and selected indicators of system operation and dispatch.

\subsubsection{Objective Function}
\label{sec:results_psom_obj}

\begin{figure*}
    \centering
    \includegraphics[width=1\linewidth]{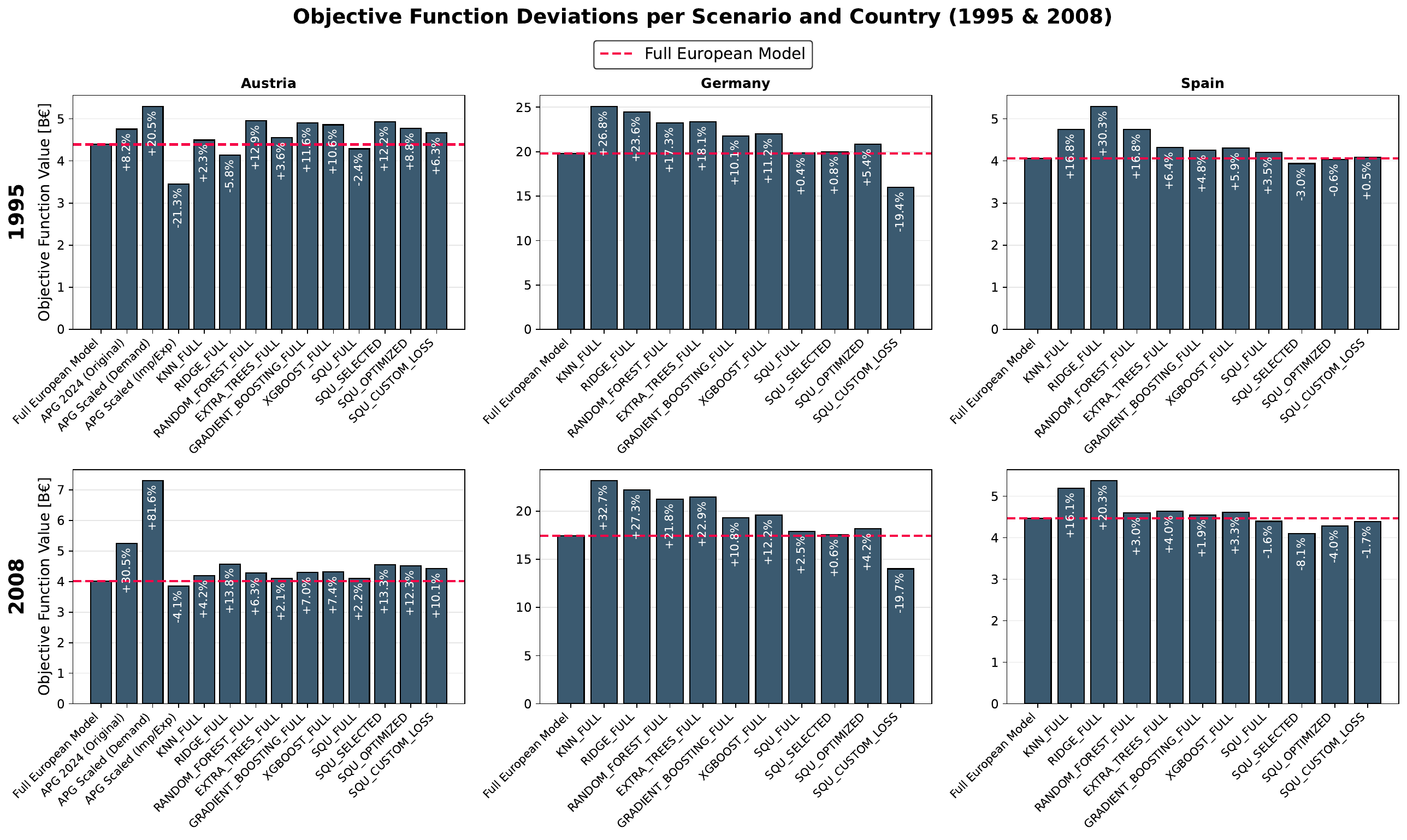}
    \caption{Total system cost for Austria (AT), Germany (DE), and Spain (ES) under alternative cross-border flow specifications for the two out-of-sample weather years 1995 and 2008. Dashed lines indicate the \texttt{Full European PSOM Model} reference.}
    \label{fig:psom_objective_function}
\end{figure*}

Figure~\ref{fig:psom_objective_function} compares the objective function values (i.e., total annual system cost)  of the single-country PSOMs against the corresponding \texttt{Full European Model} benchmark. To ensure a consistent comparison, the benchmark objective is evaluated for the same country only (i.e., using the country-specific component of the full European solution) and is shown as a dashed line. The deviations therefore quantify the economic distortion introduced when cross-border exchanges are prescribed exogenously rather than endogenously determined in the full European optimization. Results are reported for the two out-of-sample climate years (1995 and 2008).

In 1995, the APG-based benchmarks generate large cost distortions: the unscaled 2024 profile increases the objective by \SI[retain-explicit-plus=true]{+8.2}{\percent}, scaling with demand increases it to \SI[retain-explicit-plus=true]{+20.5}{\percent}, while scaling to match annual import/export totals produces a pronounced decrease of \SI{-21.3}{\percent}. In 2008, these distortions become even more pronounced (\SI[retain-explicit-plus=true]{+30.5}{\percent} for the unscaled APG series and \SI[retain-explicit-plus=true]{+81.6}{\percent} when demand-scaled), whereas scaling-to-Imp/Exp reduces the deviation to \SI{-4.1}{\percent}. This widening spread across climate years reinforces that matching annual aggregates can be misleading: the objective is driven by the temporal alignment of imports and exports with domestic residual demand and renewable availability, which directly affects hydro flexibility usage and the need for thermal dispatch.

The full-feature baseline models for Austria substantially reduce the objective distortion compared with the APG-based approaches, although their performance varies considerably. In 1995, \texttt{EXTRA\_TREES\_FULL} provides the closest conventional approximation (\SI[retain-explicit-plus=true]{+3.6}{\percent}), followed by \texttt{KNN\_FULL} (\SI[retain-explicit-plus=true]{+2.3}{\percent}), whereas \texttt{RIDGE\_FULL}, \texttt{RANDOM\_FOREST\_FULL}, \texttt{GRADIENT\_BOOSTING\_FULL}, and \texttt{XGBOOTS\_FULL} exhibit larger deviations ranging from \SI{-5.8}{\percent} to \SI[retain-explicit-plus=true]{+12.9}{\percent}. The SQU variants remain competitive, with \texttt{SQU\_FULL} slightly underestimating the objective (\SI{-2.4}{\percent}) and the remaining variants deviating between \SI[retain-explicit-plus=true]{+6.3}{\percent} and \SI[retain-explicit-plus=true]{+12.2}{\percent}. A similar picture emerges for 2008. Among the conventional baselines, \texttt{EXTRA\_TREES\_FULL} again performs best (\SI[retain-explicit-plus=true]{+2.1}{\percent}), followed by \texttt{KNN\_FULL} (\SI[retain-explicit-plus=true]{+4.2}{\percent}) and \texttt{RANDOM\_FOREST\_FULL} (\SI[retain-explicit-plus=true]{+6.3}{\percent}), while \texttt{RIDGE\_FULL}, \texttt{GRADIENT\_BOOSTING\_FULL}, and \texttt{XGBOOTS\_FULL} show larger deviations. The SQU variants span \SI[retain-explicit-plus=true]{+2.2}{\percent} to \SI[retain-explicit-plus=true]{+13.3}{\percent}. Overall, Austria demonstrates that several ML approaches substantially improve the economic consistency of the reduced PSOM compared with historical exchange profiles, while the apparent ranking of individual models remains climate-year dependent.

Germany exhibits a clearer separation between the conventional baselines and the SQU variants. In 1995, all full-feature baseline models overestimate system cost, with deviations ranging from \SI[retain-explicit-plus=true]{+10.1}{\percent} for \texttt{GRADIENT\_BOOSTING\_FULL} to \SI[retain-explicit-plus=true]{+26.8}{\percent} for \texttt{KNN\_FULL}. \texttt{GRADIENT\_BOOSTING\_FULL} and \texttt{XGBOOTS\_FULL} provide the strongest conventional performance, followed by \texttt{RANDOM\_FOREST\_FULL} and \texttt{EXTRA\_TREES\_FULL}, whereas \texttt{RIDGE\_FULL} and \texttt{KNN\_FULL} perform worst. In contrast, the standard SQU variants remain much closer to the benchmark, with deviations between \SI[retain-explicit-plus=true]{+0.4}{\percent} and \SI[retain-explicit-plus=true]{+5.4}{\percent}. The same ranking is observed in 2008. The best conventional baselines, \texttt{GRADIENT\_BOOSTING\_FULL} and \texttt{XGBOOST\_FULL}, reduce the deviation to approximately \SI[retain-explicit-plus=true]{+11}{\percent}, but remain substantially less accurate than \texttt{SQU\_FULL}, \texttt{SQU\_SELECTED}, and \texttt{SQU\_OPTIMIZED}, whose deviations range from only \SI[retain-explicit-plus=true]{+0.6}{\percent} to \SI[retain-explicit-plus=true]{+4.2}{\percent}. In contrast, \texttt{SQU\_CUSTOM\_LOSS} consistently underestimates the objective by approximately \SI{-20}{\percent} in both climate years. This systematic deviation indicates that the feasibility-oriented penalty alters the implied import--dispatch balance, favoring higher imports and therefore lower domestic generation costs than observed in the full European benchmark.

For Spain \texttt{GRADIENT\_BOOSTING\_FULL} and \texttt{XGBOOTS\_FULL} provide the closest conventional approximations for 1995, with deviations of \SI[retain-explicit-plus=true]{+4.8}{\percent} and \SI[retain-explicit-plus=true]{+5.9}{\percent}, respectively, whereas \texttt{KNN\_FULL}, \texttt{RIDGE\_FULL}, and \texttt{RANDOM\_FOREST\_FULL} overestimate the objective by approximately \SI{17}{\percent} to \SI{30}{\percent}. The SQU variants remain closest to the benchmark, ranging from \SI{-3.0}{\percent} to \SI[retain-explicit-plus=true]{+3.5}{\percent}. In 2008, the conventional tree ensembles again outperform \texttt{KNN\_FULL} and \texttt{RIDGE\_FULL}, with \texttt{EXTRA\_TREES\_FULL}, \texttt{GRADIENT\_BOOSTING\_FULL}, and \texttt{XGBOOST\_FULL} limiting deviations to between \SI[retain-explicit-plus=true]{+1.9}{\percent} and \SI[retain-explicit-plus=true]{+4.0}{\percent}. Nevertheless, the SQU variants remain closest overall, with deviations between \SI{-8.1}{\percent} and \SI{-1.6}{\percent}. These results indicate that even for a less interconnected system, accurately reproducing the timing and direction of cross-border exchanges has a measurable impact on cost-optimal system operation.

Across all three countries and both out-of-sample climate years, the SQU-based surrogates generally provide the most consistent approximation of the benchmark objective function, outperforming the conventional full-feature baselines in terms of downstream PSOM accuracy. Although Gradient Boosting and XGBoost achieve competitive prediction accuracy, particularly for Germany, the SQU variants translate these predictions more reliably into economically consistent reduced-model solutions.

\subsubsection{Generation Mix and Storage Operation}
\label{sec:results_psom_dispatch}

\begin{figure*}
    \centering
    \includegraphics[width=1\linewidth]{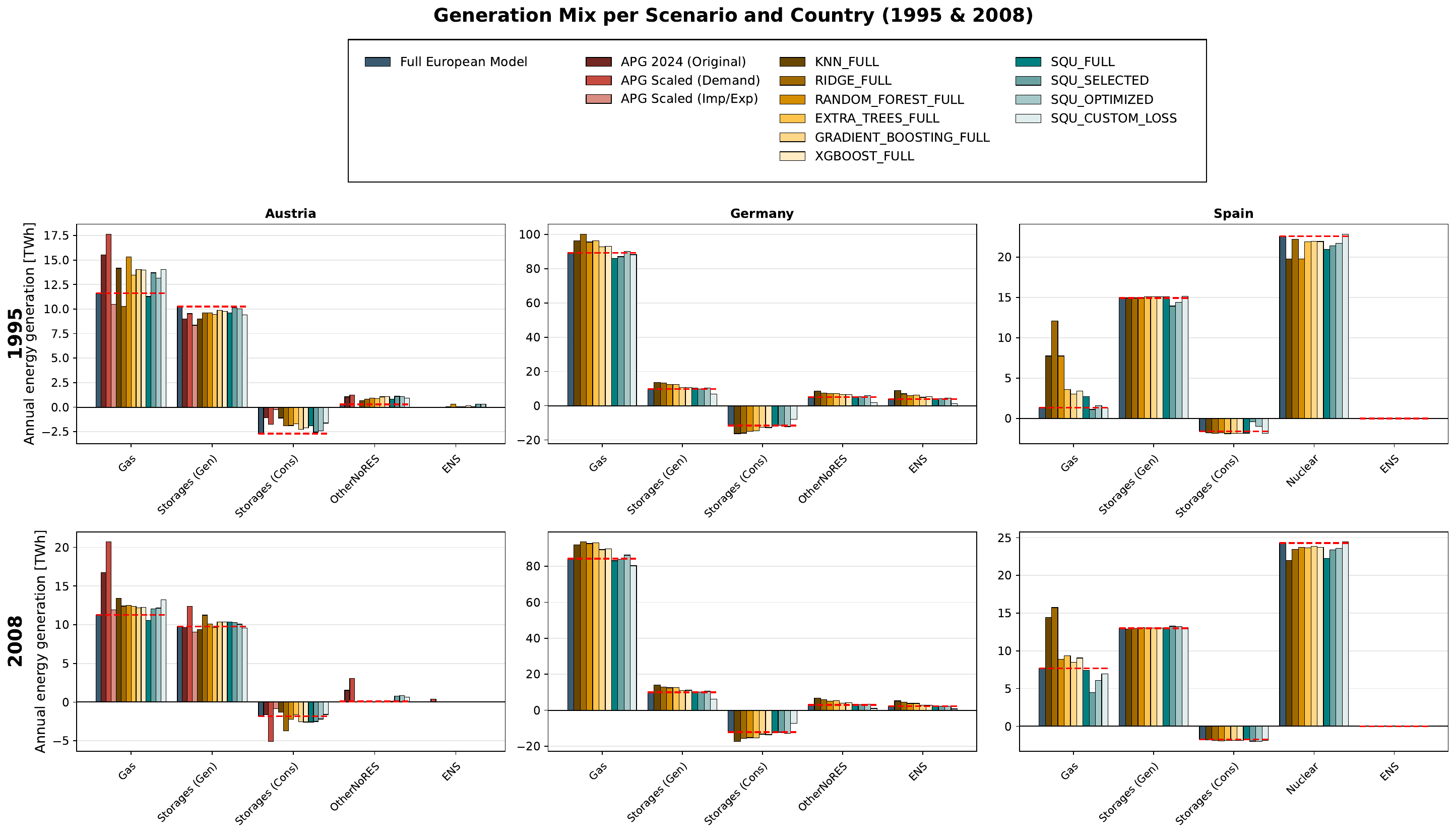}
    \caption{Annual generation of dispatchable technologies, storage generation and charging, and ENS for Austria, Germany, and Spain under alternative cross-border flow specifications for climate years~1995 and~2008. The comparison includes the APG profiles for Austria, six full-feature tabular baselines, and the four SQU variants. Storage charging is shown as negative energy; dashed lines indicate the corresponding full European PSOM reference.}
    \label{fig:psom_dispatch_dev}
\end{figure*}

While the objective function condenses the system response into a single cost figure, the dispatch decomposition reveals \emph{where} and \emph{why} alternative flow specifications affect the reduced model. Figure~\ref{fig:psom_dispatch_dev} reports annual energy volumes by dispatchable technology group (positive values denote generation, negative values denote storage charging), with dashed red segments indicating the \texttt{Full European Model} reference for each group. Non-dispatchable renewable generation (e.g., wind, PV, and run-of-river) is omitted because it is prescribed via capacity factors and thus identical across scenarios.

A key operational outcome in this comparison is ENS, because ENS directly signals infeasibilities (or scarcity) induced by the imposed interconnector-flow time series. The \texttt{SQU\_CUSTOM\_LOSS} formulation is designed to reduce such physically implausible flow patterns.

Austria’s operational response is dominated by gas-fired generation and pumped-hydro/storage operation, reflecting the country’s role as a highly interconnected system with substantial flexible capacity. In both out-of-sample years, the APG-based benchmarks produce the largest distortions in dispatch. In particular, the APG (original) and demand-scaled variants imply a substantially different balancing requirement than the benchmark, which translates into higher gas generation and altered storage cycling relative to the dashed reference markers. These effects persist across 1995 and 2008 and reinforce that transferring a historical exchange pattern from a structurally different system year can misrepresent the flexibility needed by dispatchable generators. The ML-based surrogates cluster more closely around the benchmark across gas and storage charging/discharging, indicating that they reproduce the exchange dynamics sufficiently well to preserve the intended role of Austrian hydro and storage flexibility in system balancing.
Importantly, Austria also illustrates the ENS mechanism most clearly: when the single-country PSOM is driven by ML-generated flows, a non-zero ENS appears for essentially all ML variants, indicating that small but system-relevant inconsistencies in the imposed exchange profile can translate into scarcity hours in the reduced model. By contrast, \texttt{SQU\_CUSTOM\_LOSS} eliminates ENS in Austria, consistent with its design goal of discouraging flow profiles that violate basic feasibility limits. This shows that, for Austria, feasibility-aware training improves \emph{decision relevance}: the surrogate not only matches flows statistically, but also produces boundary trajectories that avoid inducing artificial shortage in the downstream optimization.

Quantitatively, \texttt{SQU\_CUSTOM\_LOSS} reduces the artificial Austrian ENS induced by \texttt{SQU\_FULL} to zero in both CY~1995 and CY~2008, corresponding to a \SI{100}{\percent} reduction in each year. This feasibility improvement is accompanied by a modest loss in flow-prediction accuracy: the Austrian NMAE increases from \num{0.39} to \num{0.42} in CY~1995 (a relative increase of approximately \SI{7.7}{\percent}) and from \num{0.38} to \num{0.40} in CY~2008 (approximately \SI{5.3}{\percent}). The result illustrates the intended trade-off of the custom loss: a small deterioration in statistical fit can materially improve the operational feasibility of the downstream reduced model.

 an additional feasibility stress test, we applied the Austrian \texttt{SQU\_CUSTOM\_LOSS} flow predictions to the reduced Austrian LEGO model for every climate year from 1982 through 2016. Across all \num{35} climate years, annual ENS was zero. This extends the two detailed case-study years and provides evidence that the custom-loss formulation avoids surrogate-induced shortage in Austria across a wide range of historical weather realizations. It remains an empirical robustness result for the fixed 2030 model configuration, not a formal feasibility guarantee under changed capacities, prices, topology, or market rules.

Germany exhibits much larger absolute dispatch volumes (note the different axis scale), but the qualitative pattern is similar across both climate years: deviations concentrate in gas generation and storage cycling, while \textit{OtherNoRES} remain comparatively stable. The full-feature baseline models tend to induce systematically higher gas output than the benchmark, consistent with their less accurate representation of imports/exports, whereas the standard SQU variants remain closer to the reference markers in both 1995 and 2008. A notable difference to Austria is that the \texttt{Full European Model} benchmark already contains a small amount of ENS for Germany. In this setting, \texttt{SQU\_CUSTOM\_LOSS} reduces ENS even further in the single-country PSOM compared to the benchmark, which is consistent with the feasibility penalty pushing the flow profile toward states that avoid shortage.

Spain, as a more weakly interconnected system, shows smaller relative sensitivity in storage operation and nuclear output across most specifications and in both climate years. The main visible differences are concentrated in gas generation, where the full-feature baseline models deviate more strongly from the benchmark than the SQU variants. This indicates that even for a less coupled system, correctly capturing exchange timing still matters for the thermal balancing requirement, especially in periods where imports would otherwise displace domestic gas generation. Spain exhibits no ENS in these experiments, and the ENS bars remain at zero across scenarios and both out-of-sample years. Consequently, the custom-loss formulation has limited room to improve feasibility and therefore changes little in the operational outcomes relative to the other SQU variants.

Overall, the dispatch comparison reinforces two points. First, reproducing annual trade totals is not sufficient: small timing inconsistencies in the imposed interconnector flows can manifest as operational distortions and even ENS in reduced PSOMs. Second, feasibility-aware training can be valuable, but its downstream impact is system-dependent: it is clearly beneficial in Austria, where it removes surrogate-induced ENS. However, in countries where the full European PSOM already contains a small amount of ENS, the same feasibility penalty can alter the trade-off between statistical fidelity and operational feasibility and may therefore not improve overall consistency.

\subsubsection{Model Generation and Solution Times}
\label{sec:results_psom_times}

Using ML surrogates changes the PSOM runtime profile in a beneficial way. First, PSOM model generation becomes much faster because the single-country formulation contains far fewer components than the full European model. Second, PSOM solution times decrease sharply since the solver operates on a smaller and less tightly coupled optimization problem. The remaining overhead stems from the ML pipeline (data preparation, training, and prediction). For well-behaved surrogates, this overhead is small compared to the time saved in optimization. The requirement to solve the Full European Model for generating training targets does not imply that a new full-system run is required for every additional climate year. Under the same structural scenario assumptions,  including generation capacities, transmission capacities, fuel and carbon prices, and market formulation, the surrogate is trained once and can then be applied to the demand and renewable-generation profiles of additional climate years without retraining. Consequently, for a fixed structural scenario, only a single Full European Model simulation is required. Additional climate years can be evaluated using the trained surrogate without rerunning the pan-European optimization. Thus, for a robustness analysis covering, for example, \num{35} climate years, the Full European Model has to be solved only once for the training climate year, whereas the remaining \num{34} climate years require only ML prediction and reduced single-country PSOM runs. The one-time full-model and training costs are therefore amortized over the multi-year analysis. Retraining would be required only when the underlying structural scenario assumptions change materially. The speed-ups reported below compare the surrogate preparation, prediction, and reduced-PSOM run with one additional Full European Model simulation; they do not repeatedly assign the one-time full-model training-data generation cost to every surrogate application.

All surrogate-based approaches substantially reduce computational effort compared with the Full European Model, which requires approximately \SI{5906}{s} per annual simulation. Depending on the country and surrogate model, total runtimes range from approximately \SI{10}{s} to \SI{867}{s}, corresponding to speed-ups between $6.8\times$ and $582.4\times$.

Among the conventional baselines, ridge regression is consistently the fastest, achieving speed-ups of $255.2\times$, $254.7\times$, and $582.4\times$ for Austria, Germany, and Spain, respectively. KNN exhibits similarly low runtimes, while the tree-based ensemble methods incur higher one-time training costs. Random forests require approximately \SIrange{20}{22}{s} for model training, Extra Trees reduce this to about \SI{8}{s}, and XGBoost and gradient boosting exhibit the highest training costs, particularly for Germany where training requires up to \SI{89.9}{s}. Nevertheless, all tree ensembles still reduce total computation time by more than one order of magnitude relative to the Full European Model, with overall speed-ups ranging from approximately $52\times$ to $333\times$.

Within the SQU family, \texttt{SQU\_FULL} provides the most attractive runtime characteristics, achieving speed-ups between $161\times$ and $242\times$. The additional feature-selection and hyperparameter-optimization stages increase the one-time offline effort for \texttt{SQU\_SELECTED}, while \texttt{SQU\_OPTIMIZED} and \texttt{SQU\_CUSTOM\_LOSS} require approximately \SIrange{730}{840}{s} for automated hyperparameter optimization and network training, reducing the total speed-up to around $7\times$--$8\times$. However, these additional costs are incurred only once during model development.

Consequently, once a surrogate has been trained, repeated analyses for additional climate years require only the inexpensive prediction step together with the reduced PSOM solution. Since no retraining and no additional Full European Model runs are necessary.

\begin{table}[htbp]
\centering
\caption{Speed comparison of the Full European Model and single-country PSOM runs using ML flow surrogates. The ML column includes one-time surrogate preparation and training together with mean annual prediction time, but excludes the separate Full European Model run required to generate the training targets. The reported speed-ups therefore represent marginal or amortized speed-ups for repeated applications under unchanged structural scenario assumptions. For additional climate years, only prediction and reduced-PSOM solution times recur.}
\label{tab:runtime_speed_comparison}
\resizebox{\columnwidth}{!}{%
\begin{tabular}{llrrrrr}
\toprule
\textbf{Ctry.} & \textbf{Case} & \textbf{ML} &
\begin{tabular}[c]{@{}r@{}}\textbf{PSOM}\\\textbf{generation}\end{tabular} &
\begin{tabular}[c]{@{}r@{}}\textbf{PSOM}\\\textbf{solution}\end{tabular} &
\begin{tabular}[c]{@{}r@{}}\textbf{Total}\\\textbf{time}\end{tabular} &
\begin{tabular}[c]{@{}r@{}}\textbf{Speed-}\\\textbf{up}\end{tabular} \\
\hline
 & & [s] & [s] & [s] & [s] & $\times$ \\
\toprule
-- & Full European Model & \num{0} & \num{445} & \num{5462} & \num{5906} & \num{1.0} \\
\hline
AT & \texttt{KNN\_FULL} & \num{2} & \num{7} & \num{16} & \num{24} & \num{244.1} \\
AT & \texttt{RIDGE\_FULL} & \num{0.14} & \num{7} & \num{16} & \num{23.14} & \num{255.2} \\
AT & \texttt{RANDOM\_FOREST\_FULL} & \num{20.25} & \num{7} & \num{16} & \num{43.25} & \num{136.6} \\
AT & \texttt{EXTRA\_TREES\_FULL} & \num{7.76} & \num{7} & \num{16} & \num{30.76} & \num{192.0} \\
AT & \texttt{GRADIENT\_BOOSTING\_FULL} & \num{25.71} & \num{7} & \num{16} & \num{48.71} & \num{121.3} \\
AT & \texttt{XGBOOST\_FULL} & \num{19.42} & \num{7} & \num{16} & \num{42.42} & \num{139.2} \\
AT & \texttt{SQU\_FULL} & \num{13} & \num{7} & \num{11} & \num{32} & \num{186.0} \\
AT & \texttt{SQU\_SELECTED} & \num{38} & \num{7} & \num{16} & \num{61} & \num{96.7} \\
AT & \texttt{SQU\_OPTIMIZED} & \num{775} & \num{7} & \num{11} & \num{792} & \num{7.5} \\
AT & \texttt{SQU\_CUSTOM\_LOSS} & \num{786} & \num{7} & \num{16} & \num{810} & \num{7.3} \\
\hline
DE & \texttt{KNN\_FULL} & \num{4} & \num{9} & \num{14} & \num{27} & \num{215.1} \\
DE & \texttt{RIDGE\_FULL} & \num{0.18} & \num{9} & \num{14} & \num{23.18} & \num{254.7} \\
DE & \texttt{RANDOM\_FOREST\_FULL} & \num{21.48} & \num{9} & \num{14} & \num{44.48} & \num{132.8} \\
DE & \texttt{EXTRA\_TREES\_FULL} & \num{8.11} & \num{9} & \num{14} & \num{31.11} & \num{189.8} \\
DE & \texttt{GRADIENT\_BOOSTING\_FULL} & \num{89.92} & \num{9} & \num{14} & \num{112.92} & \num{52.3} \\
DE & \texttt{XGBOOST\_FULL} & \num{54.84} & \num{9} & \num{14} & \num{77.84} & \num{75.9} \\
DE & \texttt{SQU\_FULL} & \num{14} & \num{8} & \num{15} & \num{37} & \num{160.9} \\
DE & \texttt{SQU\_SELECTED} & \num{94} & \num{8} & \num{15} & \num{117} & \num{50.4} \\
DE & \texttt{SQU\_OPTIMIZED} & \num{832} & \num{8} & \num{14} & \num{855} & \num{6.9} \\
DE & \texttt{SQU\_CUSTOM\_LOSS} & \num{844} & \num{9} & \num{15} & \num{867} & \num{6.8} \\
\hline
ES & \texttt{KNN\_FULL} & \num{1} & \num{3} & \num{7} & \num{11} & \num{525.5} \\
ES & \texttt{RIDGE\_FULL} & \num{0.14} & \num{3} & \num{7} & \num{10.14} & \num{582.4} \\
ES & \texttt{RANDOM\_FOREST\_FULL} & \num{21.57} & \num{3} & \num{7} & \num{31.57} & \num{187.1} \\
ES & \texttt{EXTRA\_TREES\_FULL} & \num{7.72} & \num{3} & \num{7} & \num{17.72} & \num{333.3} \\
ES & \texttt{GRADIENT\_BOOSTING\_FULL} & \num{9.37} & \num{3} & \num{7} & \num{19.37} & \num{304.8} \\
ES & \texttt{XGBOOST\_FULL} & \num{9.08} & \num{3} & \num{7} & \num{19.08} & \num{309.5} \\
ES & \texttt{SQU\_FULL} & \num{14} & \num{3} & \num{7} & \num{24} & \num{242.2} \\
ES & \texttt{SQU\_SELECTED} & \num{71} & \num{3} & \num{7} & \num{82} & \num{72.2} \\
ES & \texttt{SQU\_OPTIMIZED} & \num{728} & \num{3} & \num{7} & \num{739} & \num{8.0} \\
ES & \texttt{SQU\_CUSTOM\_LOSS} & \num{737} & \num{3} & \num{7} & \num{747} & \num{7.9} \\
\bottomrule
\end{tabular}
}
\end{table}

\section{Discussion}
\label{sec:Discussion}

\added{The core contribution of this work should not be interpreted as a new generic regression algorithm. Rather, it is the use of learned interconnector-level flows as transferable boundary information between a computationally expensive full-system PSOM and repeated geographically reduced PSOM studies. This distinction is practically relevant whenever the research question is regional but the region remains strongly coupled to the surrounding power system. The literature comparison in Table~\ref{tab:literature_comparison} shows that previous studies address individual elements of this problem, such as cross-border-flow prediction, external-system representation, or ML approximation of the original OPF, whereas the present framework combines full-PSOM-generated flow targets, learned boundary conditions, and downstream reduced-PSOM validation.}

\added{The intended scope is also important when interpreting feasibility. The surrogate is not proposed for real-time power-system operation, security-constrained dispatch, or detailed transmission- or distribution-grid planning, for which a feasible power-flow solution and the relevant network constraints are mandatory. Instead, the framework targets energy-system planning and scenario studies in which the principal quantities of interest include generation mixes, storage operation, system costs, curtailment, ENS, and energy exchange with neighboring systems. In such applications, geographically reduced models can provide substantial computational advantages, provided that the influence of the external system is represented sufficiently well for the research question under investigation.}

\added{Accordingly, the proposed framework should be interpreted as a computational model-reduction and screening approach rather than a feasibility-guaranteed replacement for the full PSOM. Predicted interconnector flows are evaluated by solving the downstream reduced PSOM. Cases exhibiting surrogate-induced ENS or otherwise unacceptable behavior can subsequently be evaluated with the full European model. This workflow is particularly attractive for multi-climate-year or large-sensitivity studies, because the computationally expensive full-system model needs to be revisited only for cases for which the surrogate representation proves inadequate.}

The results show that the choice of cross-border flow representation is not a minor modeling detail, but a structural assumption that can materially affect both statistical fit and optimization outcomes. In particular, the comparison against APG-based profiles highlights that ``fixed-flow'' practices can fail even when they appear reasonable at an aggregate level. For Austria, the 2024 APG series and its scaled variants deliver strongly negative $R^2$ values in both out-of-sample climate years, indicating that they are not only inaccurate but systematically inconsistent with the flow patterns implied by the 2030 system assumptions. The key issue is not the annual net balance per se, but the intra-annual timing of imports and exports relative to residual demand and renewable availability. Scaling can force annual totals to align by construction, yet still produces flow shapes that shift scarcity hours, storage cycling, and thermal dispatch in ways that distort the objective value and operational indicators. This reinforces a central point: for reduced PSOMs, the temporal structure of cross-border exchanges is decision-relevant.

However, the surrogate does not learn the real European power system. It learns the flow behavior of a specific simplified pan-European PSOM DC-OPF formulation under a fixed scenario, climate-year set, and solver configuration. The resulting flow profiles are therefore model-consistent synthetic boundary conditions, not forecasts of actual future market outcomes.

A key implication is that statistical improvements translate into more consistent optimization outcomes when the flows are used as fixed boundary conditions. In the single-country PSOMs, SQU-based flow specifications cluster closer to the corresponding country components of the full European benchmark in terms of total system cost, annual trade balances, and dispatch patterns. This indicates that the surrogate is not merely fitting point-wise errors, but preserving enough of the timing and sign structure of exchanges to maintain the role of domestic flexibility resources. Austria illustrates this mechanism clearly: deviations in annual imports are typically compensated by gas generation and storage operation, so mis-specified flow timing quickly propagates into distorted dispatch. Germany shows smaller relative sensitivity in many cases due to system scale and portfolio diversity, yet even modest percentage shifts remain economically meaningful. Spain is less driven by cross-border exchanges in magnitude, but improved flow timing still affects the residual demand met by dispatchable units and hence the cost outcome.

Feasibility-aware training provides an additional, operationally meaningful lever. The custom-loss formulation augments the prediction error with a penalty for physically implausible net-export profiles, thereby targeting surrogate-induced infeasibilities that manifest as ENS once the flows are imposed in the reduced PSOM. The results indicate that this mechanism can work as intended when the benchmark itself is feasible: in Austria, where the full European benchmark exhibits no ENS, several ML-based flow specifications induce small surrogate-driven ENS, whereas \texttt{SQU\_CUSTOM\_LOSS} removes these artifacts and improves operational consistency. The additional robustness check, where the Austrian \texttt{SQU\_CUSTOM\_LOSS} surrogate was evaluated for every available climate year from 1982 to 2016 shows, that across all 35 climate years, the reduced PSOM remained free of artificial ENS, indicating that the feasibility-aware penalty generalizes well across weather variability under fixed structural assumptions. For Germany, where the full European benchmark contains a small amount of ENS, the custom loss can reduce ENS further in the reduced model, but this comes with the risk of shifting imports and dispatch away from the benchmark trajectory and thereby affecting the objective. In Spain, ENS does not occur, and the custom loss has limited impact. Overall, feasibility-aware training is most effective for eliminating surrogate-induced infeasibilities; its benefits are system-dependent and may require adaptive designs to avoid unintended economic biases when the benchmark itself contains scarcity.

A first methodological limitation concerns uniqueness. In linear OPF and energy-system models with aggregated technologies, identical exogenous demand and renewable-generation profiles may admit several cost-equivalent or near-equivalent dispatch solutions, which can imply different interconnector flows. The supervised target is therefore the solution produced by a specific full PSOM instance and solver configuration. The surrogate learns the deterministic mapping induced by the formulation, data, solver, numerical tolerances, and tie-breaking behavior rather than a unique physical law. This is appropriate when the objective is to reproduce a particular modeling workflow.

A second limitation concerns generalization. The surrogate implicitly learns the dispatch and net-injection response of the full PSOM under fixed structural assumptions. The out-of-sample tests on CY~1995 and CY~2008 therefore evaluate robustness to unseen weather realizations, but not to changes in fuel prices, carbon prices, installed generation or storage capacities, market rules, transmission capacities, network topology, or bidding-zone configurations. Because such changes alter the hidden mapping from exogenous inputs to dispatch, net injections, and flows, transfer to substantially different system conditions would require new full-model simulations followed by retraining the ML models. Future work could therefore investigate scenario-conditioned or transfer-learning surrogates that remain applicable across multiple structural scenarios.

\added{A third limitation concerns feasibility and optimality. Feasibility is indispensable when ML outputs are intended to replace physics-based operating or security calculations. That is not the role claimed for the present framework. The custom loss can reduce net-export predictions that exceed an approximate domestic supply bound and thereby reduce surrogate-induced ENS, but it remains a soft penalty: it is neither a DC-OPF feasibility projection nor a guarantee of reduced-model or full-system feasibility. This limitation is why downstream PSOM feasibility and ENS are explicitly evaluated and why problematic scenarios can be returned to the full European model.}

Relative to decomposition-based and learning-assisted optimization methods, the proposed approach intentionally makes a different trade-off. Decomposition approaches preserve an explicit link to the original optimization problem and can provide convergence or optimality guarantees, but they generally require iterative coordination or repeated subproblem solves. Our surrogate workflow is non-iterative once trained and is designed for repeated reduced-model studies where a fully coupled European solve is impractical. It therefore improves practical tractability and boundary-condition consistency, but does not solve the original full OPF more efficiently or provide formal guarantees.

The runtime results should therefore be interpreted in an amortized sense. The full-model solve needed to generate training data, feature selection, hyperparameter tuning, and model training must be counted as preparation costs. The approach becomes attractive when these costs are reused over many reduced-model runs, for example across climate years.

Overall, the framework is most appropriate when (i) a full-system PSOM can be solved for a representative training scenario, (ii) many reduced-model studies need consistent cross-border boundary conditions, and (iii) the target scenarios remain within or near the distribution represented by the training data. It is less appropriate when formal feasibility guarantees, market-clearing guarantees, or extrapolation to structurally different systems are required.

\section{Conclusions and Outlook}
\label{sec:ConclusionOutlook}

\added{The core contribution of this work is an ML surrogate framework for transferring cross-border interactions from a full-system PSOM to geographically reduced PSOM studies through predicted interconnector-level boundary flows. The surrogate learns cross-border flows generated by the full European PSOM and provides them as fixed boundary conditions to reduced models, thereby avoiding repeated full-European optimization runs when multiple climate years are analyzed under unchanged structural assumptions.} The framework was implemented with the open-source LEGO DC-OPF model under ENTSO-E TYNDP~2024 \emph{National Trends} assumptions for 2030. Surrogates were trained on climate year~2009 and evaluated on two fully out-of-sample climate years (1995 and 2008) for Austria, Germany, and Spain.

The results show, within the considered LEGO DC-OPF setup, that commonly used fixed-flow approximations based on historical import/export profiles can become highly inconsistent under changing weather realizations and future system configurations. In the Austrian case, APG~2024 baselines (including scaled variants) perform poorly in both out-of-sample years and can induce substantial distortions in reduced-model outcomes. By contrast, ML-based surrogates provide robust improvements across countries and both out-of-sample years. In particular, the neural-network family (SQU) generalizes more robustly than the conventional full-feature baselines to the out-of-sample climate years and yields single-country PSOM outcomes that are markedly closer to the corresponding country components of the full European benchmark in total system cost, annual trade balances, and dispatch patterns. This indicates that, under fixed scenario assumptions, the surrogates capture part of the nonlinear dependence of exchanges on renewable availability and demand sufficiently well to preserve the role of domestic flexibility resources in reduced optimization models.

\added{As an additional contribution, the feasibility-aware custom loss softly penalizes net-export profiles that exceed an approximate domestic supply bound. It does not guarantee complete power-flow feasibility, but it is effective at suppressing \emph{surrogate-induced} ENS in the examined application:} in Austria, where the full European benchmark exhibits no ENS, the custom loss eliminates ENS in the reduced single-country runs and thereby improves operational consistency. This result is further supported by the Austrian climate-year stress test for 1982--2016, in which the reduced LEGO model produced zero ENS for all \num{35} years when driven by \texttt{SQU\_CUSTOM\_LOSS} flows.

From a computational perspective, the workflow offers large speedups for repeated reduced-model solves because reduced single-country PSOMs are substantially smaller than the full European model and surrogate prediction is inexpensive once trained. In addition, the approach is naturally parallelizable: countries and climate years can be evaluated independently, enabling efficient multi-country and multi-year analyses.

Building on these results, a key next step is to increase spatial granularity by predicting flows per interconnector or transmission line in a fully European setting with an explicit high-voltage grid representation, allowing reduced models to reflect congestion and regional trade patterns more realistically. Further work should assess the approach in generation expansion planning models, broaden the surrogate output space (e.g., zonal prices, congestion indicators, reserve requirements), and develop probabilistic surrogates to quantify uncertainty and propagate it into planning decisions. Another promising direction is decision-focused learning, in which the surrogate is trained directly with respect to downstream optimization quality rather than prediction accuracy alone. Finally, feasibility-aware training warrants deeper investigation, including hybrid or country-specific constraints that reduce ENS without introducing systematic biases in implied trade and dispatch patterns. Future work includes extending the feasibility-aware loss toward more comprehensive physics-informed surrogate models that embed additional network and operational constraints directly into the learning process.

\added{The framework does not guarantee power-flow feasibility and is not intended to replace physics-based operational or security analyses. Instead, the reduced PSOM serves as a downstream consistency check, with the full European model available when the surrogate representation is insufficient. Future work could build on advances in physics-informed learning and feasibility-restoring methods to extend the current soft constraint toward stricter or hard feasibility enforcement.}

Overall, the proposed framework demonstrates that learning boundary conditions, rather than complete OPF solutions, provides a practical and computationally efficient pathway for geographically reduced power system optimization under fixed structural scenarios.

\section*{Acknowledgements}
\label{Acknowledgements}

R. Gaugl gratefully acknowledges the Institute for Research in Technology (IIT) – School of Engineering (ICAI) at the Universidad Pontificia Comillas for hosting his research visit and the funding granted by the Rudolf Chaudoire Foundation and the Erasmus+ program.

\section*{CRediT authorship contribution statement}
\textbf{Robert~Gaugl}: Writing -- original draft, Conceptualization, Methodology, Software, Formal analysis, Visualization. 
\textbf{Eloy~Insunza~Díaz}: Writing -- review \& editing, Methodology, Validation. 
\textbf{José~Portela~González}: Writing -- review \& editing, Supervision, Methodology, Validation. 
\textbf{Sonja~Wogrin}: Writing -- review \& editing, Supervision, Validation.

\section*{Disclaimer}
\label{Disclaimer}

The authors acknowledge the use of OpenAI's ChatGPT (version GPT-5.2, February 2026) for assistance in improving the grammar, spelling, and clarity of the manuscript. All content and conclusions are the sole responsibility of the authors.




\bibliographystyle{elsarticle-num-names.bst} 
\bibliography{references}

@inproceedings{donti_dc3_2021,
	title = {{DC3}: {A} {Learning} {Method} for {Optimization} with {Hard} {Constraints}},
	booktitle = {International {Conference} on {Learning} {Representations}},
	author = {Donti, Priya L. and Rolnick, David and Kolter, J. Zico},
	year = {2021},
}

@inproceedings{zamzam_learning_2020,
	title = {Learning {Optimal} {Solutions} for {Extremely} {Fast} {AC} {Optimal} {Power} {Flow}},
	url = {https://ieeexplore.ieee.org/document/9303008},
	doi = {10.1109/SmartGridComm47815.2020.9303008},
	urldate = {2026-09-11},
	booktitle = {2020 {IEEE} {International} {Conference} on {Communications}, {Control}, and {Computing} {Technologies} for {Smart} {Grids} ({SmartGridComm})},
	author = {Zamzam, Ahmed S. and Baker, Kyri},
	month = nov,
	year = {2020},
	pages = {1--6},
}

@article{fioretto_predicting_2020,
	title = {Predicting {AC} {Optimal} {Power} {Flows}: {Combining} {Deep} {Learning} and {Lagrangian} {Dual} {Methods}},
	volume = {34},
	copyright = {Copyright (c) 2020 Association for the Advancement of Artificial Intelligence},
	issn = {2374-3468},
	shorttitle = {Predicting {AC} {Optimal} {Power} {Flows}},
	url = {https://ojs.aaai.org/index.php/AAAI/article/view/5403},
	doi = {10.1609/aaai.v34i01.5403},
	language = {en},
	number = {01},
	urldate = {2026-09-11},
	journal = {Proceedings of the AAAI Conference on Artificial Intelligence},
	author = {Fioretto, Ferdinando and Mak, Terrence W. K. and Hentenryck, Pascal Van},
	month = apr,
	year = {2020},
	pages = {630--637},
}

@article{karniadakis_physics-informed_2021,
	title = {Physics-informed machine learning},
	volume = {3},
	copyright = {2021 Springer Nature Limited},
	issn = {2522-5820},
	url = {https://www.nature.com/articles/s42254-021-00314-5},
	doi = {10.1038/s42254-021-00314-5},
	language = {en},
	number = {6},
	urldate = {2026-07-28},
	journal = {Nature Reviews Physics},
	publisher = {Nature Publishing Group},
	author = {Karniadakis, George Em and Kevrekidis, Ioannis G. and Lu, Lu and Perdikaris, Paris and Wang, Sifan and Yang, Liu},
	month = jun,
	year = {2021},
	pages = {422--440},
}

@article{raissi_physics-informed_2019,
	title = {Physics-informed neural networks: {A} deep learning framework for solving forward and inverse problems involving nonlinear partial differential equations},
	volume = {378},
	issn = {0021-9991},
	shorttitle = {Physics-informed neural networks},
	url = {https://www.sciencedirect.com/science/article/pii/S0021999118307125},
	doi = {10.1016/j.jcp.2018.10.045},
	urldate = {2026-07-28},
	journal = {Journal of Computational Physics},
	author = {Raissi, M. and Perdikaris, P. and Karniadakis, G. E.},
	month = feb,
	year = {2019},
	pages = {686--707},
}

@misc{contributors_xgboost_2025,
	title = {{XGBoost} {Python} {Package}},
	url = {https://xgboost.readthedocs.io/},
	author = {Contributors, XGBoost},
	year = {2025},
}

@inproceedings{wilder_melding_2019,
	title = {Melding the {Data}-{Decisions} {Pipeline}: {Decision}-{Focused} {Learning} for {Combinatorial} {Optimization}},
	volume = {33},
	doi = {10.1609/aaai.v33i01.33011658},
	number = {01},
	booktitle = {Proceedings of the {AAAI} {Conference} on {Artificial} {Intelligence}},
	author = {Wilder, Bryan and Dilkina, Bistra and Tambe, Milind},
	year = {2019},
	pages = {1658--1665},
}

@article{elmachtoub_smart_2022,
	title = {Smart “{Predict}, then {Optimize}”},
	volume = {68},
	doi = {10.1287/mnsc.2020.3922},
	number = {1},
	journal = {Management Science},
	author = {Elmachtoub, Adam N. and Grigas, Paul},
	year = {2022},
	pages = {9--26},
}

@article{khaloie_review_2025,
	title = {Review of machine learning techniques for optimal power flow},
	volume = {388},
	issn = {0306-2619},
	url = {https://www.sciencedirect.com/science/article/pii/S0306261925003678},
	doi = {10.1016/j.apenergy.2025.125637},
	urldate = {2026-07-24},
	journal = {Applied Energy},
	author = {Khaloie, Hooman and Dolányi, Mihály and Toubeau, Jean-François and Vallée, François},
	month = jun,
	year = {2025},
	pages = {125637},
}

@misc{chollet_keras_2015,
	title = {Keras},
	url = {https://keras.io},
	author = {Chollet, François and {et al.}},
	year = {2015},
}

@article{eren_comprehensive_2024,
	title = {A comprehensive review on deep learning approaches for short-term load forecasting},
	volume = {189},
	issn = {1364-0321},
	url = {https://www.sciencedirect.com/science/article/pii/S1364032123008894},
	doi = {10.1016/j.rser.2023.114031},
	urldate = {2026-04-07},
	journal = {Renewable and Sustainable Energy Reviews},
	author = {Eren, Yavuz and Küçükdemiral, Ibrahim},
	month = jan,
	year = {2024},
	pages = {114031},
}

@article{bille_forecasting_2023,
	title = {Forecasting electricity prices with expert, linear, and nonlinear models},
	volume = {39},
	issn = {0169-2070},
	url = {https://www.sciencedirect.com/science/article/pii/S0169207022000036},
	doi = {10.1016/j.ijforecast.2022.01.003},
	number = {2},
	urldate = {2026-04-07},
	journal = {International Journal of Forecasting},
	author = {Billé, Anna Gloria and Gianfreda, Angelica and Del Grosso, Filippo and Ravazzolo, Francesco},
	month = apr,
	year = {2023},
	pages = {570--586},
}

@article{lago_forecasting_2021,
	title = {Forecasting day-ahead electricity prices: {A} review of state-of-the-art algorithms, best practices and an open-access benchmark},
	volume = {293},
	issn = {0306-2619},
	shorttitle = {Forecasting day-ahead electricity prices},
	url = {https://www.sciencedirect.com/science/article/pii/S0306261921004529},
	doi = {10.1016/j.apenergy.2021.116983},
	urldate = {2026-04-07},
	journal = {Applied Energy},
	author = {Lago, Jesus and Marcjasz, Grzegorz and De Schutter, Bart and Weron, Rafał},
	month = jul,
	year = {2021},
	pages = {116983},
}

@article{nowotarski_recent_2018,
	title = {Recent advances in electricity price forecasting: {A} review of probabilistic forecasting},
	volume = {81},
	issn = {1364-0321},
	shorttitle = {Recent advances in electricity price forecasting},
	url = {https://www.sciencedirect.com/science/article/pii/S1364032117308808},
	doi = {10.1016/j.rser.2017.05.234},
	urldate = {2026-04-07},
	journal = {Renewable and Sustainable Energy Reviews},
	author = {Nowotarski, Jakub and Weron, Rafał},
	month = jan,
	year = {2018},
	pages = {1548--1568},
}

@article{biswal_review_2024,
	title = {Review on smart grid load forecasting for smart energy management using machine learning and deep learning techniques},
	volume = {12},
	issn = {2352-4847},
	url = {https://www.sciencedirect.com/science/article/pii/S2352484724006346},
	doi = {10.1016/j.egyr.2024.09.056},
	urldate = {2026-04-07},
	journal = {Energy Reports},
	author = {Biswal, Biswajit and Deb, Subhasish and Datta, Subir and Ustun, Taha Selim and Cali, Umit},
	month = dec,
	year = {2024},
	pages = {3654--3670},
}

@article{hong_probabilistic_2016,
	title = {Probabilistic electric load forecasting: {A} tutorial review},
	volume = {32},
	issn = {0169-2070},
	shorttitle = {Probabilistic electric load forecasting},
	url = {https://www.sciencedirect.com/science/article/pii/S0169207015001508},
	doi = {10.1016/j.ijforecast.2015.11.011},
	number = {3},
	urldate = {2026-04-07},
	journal = {International Journal of Forecasting},
	author = {Hong, Tao and Fan, Shu},
	month = jul,
	year = {2016},
	pages = {914--938},
}

@article{pedregosa_scikit-learn_2011,
	title = {Scikit-learn: {Machine} {Learning} in {Python}},
	volume = {12},
	journal = {Journal of Machine Learning Research},
	author = {Pedregosa, F. and Varoquaux, G. and Gramfort, A. and Michel, V. and Thirion, B. and Grisel, O. and Blondel, M. and Prettenhofer, P. and Weiss, R. and Dubourg, V. and Vanderplas, J. and Passos, A. and Cournapeau, D. and Brucher, M. and Perrot, M. and Duchesnay, E.},
	year = {2011},
	pages = {2825--2830},
}

@article{wogrin_assessing_2020,
	title = {Assessing the impact of inertia and reactive power constraints in generation expansion planning},
	volume = {280},
	issn = {0306-2619},
	doi = {10.1016/J.APENERGY.2020.115925},
	urldate = {2021-08-30},
	journal = {Applied Energy},
	publisher = {Elsevier},
	author = {Wogrin, S. and Tejada-Arango, D. and Delikaraoglou, S. and Botterud, A.},
	month = dec,
	year = {2020},
	pages = {115925},
}

@article{mertens_representing_2020,
	title = {Representing cross-border trade of electricity in long-term energy-system optimization models with a limited geographical scope},
	volume = {261},
	issn = {0306-2619},
	url = {https://www.sciencedirect.com/science/article/pii/S030626191932063X},
	doi = {10.1016/j.apenergy.2019.114376},
	urldate = {2026-03-03},
	journal = {Applied Energy},
	author = {Mertens, Tim and Poncelet, Kris and Duerinck, Jan and Delarue, Erik},
	month = mar,
	year = {2020},
	pages = {114376},
}

@misc{entso-e_tyndp_2024,
	title = {{TYNDP} 2024 -{Implementation} {Guidelines}},
	url = {https://tyndp.entsoe.eu/resources/tyndp-2024-implementation-guidelines-march},
	language = {en},
	urldate = {2025-03-02},
	author = {{ENTSO-E}},
	year = {2024},
}

@misc{brown_pypsa-eur_2026,
	title = {{PyPSA}-{Eur}: {An} {Open} {Sector}-{Coupled} {Optimisation} {Model} of the {European} {Energy} {System}},
	url = {https://github.com/pypsa/pypsa-eur},
	publisher = {GitHub},
	author = {Brown, Tom and Victoria, Marta and Zeyen, Elisabeth and Hofmann, Fabian and Neumann, Fabian and Frysztacki, Martha and Hampp, Johannes and Schlachtberger, David and Hörsch, Jonas and Schledorn, Amos and Schauß, Caspar and van Greevenbroek, Koen and Millinger, Markus and Glaum, Philipp and Xiong, Bobby and Seibold, Toni},
	year = {2026},
}

@article{pizarroso_neuralsens_2022,
	title = {{NeuralSens}: {Sensitivity} {Analysis} of {Neural} {Networks}},
	volume = {102},
	copyright = {Copyright (c) 2022 Jaime Pizarroso, José Portela, Antonio Muñoz},
	issn = {1548-7660},
	shorttitle = {{NeuralSens}},
	url = {https://doi.org/10.18637/jss.v102.i07},
	doi = {10.18637/jss.v102.i07},
	language = {en},
	urldate = {2026-02-22},
	journal = {Journal of Statistical Software},
	author = {Pizarroso, Jaime and Portela, José and Muñoz, Antonio},
	month = apr,
	year = {2022},
	pages = {1--36},
}

@article{aravena_renewable_2017,
	title = {Renewable {Energy} {Integration} in {Zonal} {Markets}},
	volume = {32},
	issn = {1558-0679},
	url = {https://ieeexplore.ieee.org/document/7501844},
	doi = {10.1109/TPWRS.2016.2585222},
	number = {2},
	urldate = {2026-02-22},
	journal = {IEEE Transactions on Power Systems},
	author = {Aravena, Ignacio and Papavasiliou, Anthony},
	month = mar,
	year = {2017},
	pages = {1334--1349},
}

@article{frey_tackling_2024,
	title = {Tackling the multitude of uncertainties in energy systems analysis by model coupling and high-performance computing},
	volume = {3},
	issn = {2813-2823},
	url = {https://www.frontiersin.org/journals/environmental-economics/articles/10.3389/frevc.2024.1398358/full},
	doi = {10.3389/frevc.2024.1398358},
	language = {English},
	urldate = {2026-02-22},
	journal = {Frontiers in Environmental Economics},
	publisher = {Frontiers},
	author = {Frey, Ulrich J. and Sasanpour, Shima and Breuer, Thomas and Buschmann, Jan and Cao, Karl-Kiên},
	month = oct,
	year = {2024},
}

@article{unger_effect_2018,
	title = {The effect of wind energy production on cross-border electricity pricing: {The} case of western {Denmark} in the {Nord} {Pool} market},
	volume = {58},
	issn = {0313-5926},
	shorttitle = {The effect of wind energy production on cross-border electricity pricing},
	url = {https://www.sciencedirect.com/science/article/pii/S0313592617302011},
	doi = {10.1016/j.eap.2018.01.006},
	urldate = {2026-02-22},
	journal = {Economic Analysis and Policy},
	author = {Unger, Elizabeth A. and Ulfarsson, Gudmundur F. and Gardarsson, Sigurdur M. and Matthiasson, Thorolfur},
	month = jun,
	year = {2018},
	pages = {121--130},
}

@article{crozier_effect_2022,
	title = {The effect of renewable electricity generation on the value of cross-border interconnection},
	volume = {324},
	issn = {0306-2619},
	url = {https://www.sciencedirect.com/science/article/pii/S030626192201008X},
	doi = {10.1016/j.apenergy.2022.119717},
	urldate = {2026-02-22},
	journal = {Applied Energy},
	author = {Crozier, Constance and Baker, Kyri},
	month = oct,
	year = {2022},
	pages = {119717},
}

@techreport{entso-e_tyndp_2025,
	title = {{TYNDP} 2024 - {Scenario} {Report}},
	url = {https://2024.entsos-tyndp-scenarios.eu/},
	urldate = {2025-07-24},
	institution = {ENTSO-E},
	author = {{ENTSO-E}},
	year = {2025},
}

@article{gajowniczek_two-stage_2017,
	title = {Two-{Stage} {Electricity} {Demand} {Modeling} {Using} {Machine} {Learning} {Algorithms}},
	volume = {10},
	copyright = {http://creativecommons.org/licenses/by/3.0/},
	issn = {1996-1073},
	url = {https://www.mdpi.com/1996-1073/10/10/1547},
	doi = {10.3390/en10101547},
	language = {en},
	number = {10},
	urldate = {2025-07-23},
	journal = {Energies},
	publisher = {Multidisciplinary Digital Publishing Institute},
	author = {Gajowniczek, Krzysztof and Zabkowski, Tomasz},
	month = oct,
	year = {2017},
	note = {Number: 10},
	pages = {1547},
}

@misc{european_union_regulation_2019,
	title = {Regulation ({EU}) 2019/943 of the {European} {Parliament} and of the {Council} of 5 {June} 2019 on the internal market for electricity},
	url = {https://eur-lex.europa.eu/legal-content/EN/TXT/?uri=CELEX:32019R0943},
	author = {{European Union}},
	year = {2019},
}

@phdthesis{kledzik_predicting_2024,
	title = {Predicting cross-border power flow using weather data},
	url = {https://www.diva-portal.org/smash/record.jsf?pid=diva2%3A1946084&dswid=4245},
	language = {eng},
	urldate = {2025-07-24},
	school = {KTH Royal Institute of Technology},
	author = {Kledzik, Vilgot and Haker, Jonas},
	year = {2024},
}

@article{huang_intelligent_2023,
	title = {An {Intelligent} {Algorithm} for {Solving} {Unit} {Commitments} {Based} on {Deep} {Reinforcement} {Learning}},
	volume = {15},
	copyright = {http://creativecommons.org/licenses/by/3.0/},
	issn = {2071-1050},
	url = {https://www.mdpi.com/2071-1050/15/14/11084},
	doi = {10.3390/su151411084},
	language = {en},
	number = {14},
	urldate = {2025-07-24},
	journal = {Sustainability},
	publisher = {Multidisciplinary Digital Publishing Institute},
	author = {Huang, Guanglei and Mao, Tian and Zhang, Bin and Cheng, Renli and Ou, Mingyu},
	month = jan,
	year = {2023},
	note = {Number: 14},
	pages = {11084},
}

@article{pineda_is_2022,
	title = {Is learning for the unit commitment problem a low-hanging fruit?},
	volume = {207},
	issn = {0378-7796},
	url = {https://www.sciencedirect.com/science/article/pii/S0378779622000815},
	doi = {10.1016/j.epsr.2022.107851},
	urldate = {2025-07-24},
	journal = {Electric Power Systems Research},
	author = {Pineda, S. and Morales, J. M.},
	month = jun,
	year = {2022},
	pages = {107851},
}

@article{zhu_reinforcement_2025,
	title = {A {Reinforcement} {Learning} {Embedded} {Surrogate} {Lagrangian} {Relaxation} {Method} for {Fast} {Solving} {Unit} {Commitment} {Problems}},
	issn = {1558-0679},
	url = {https://ieeexplore.ieee.org/abstract/document/10847790},
	doi = {10.1109/TPWRS.2025.3529700},
	urldate = {2025-07-24},
	journal = {IEEE Transactions on Power Systems},
	author = {Zhu, Yuhang and Cui, Gaochen and Liu, Anbang and Jia, Qing-Shan and Guan, Xiaohong and Zhai, Qiaozhu and Guo, Qi and Guo, Xianping},
	year = {2025},
	pages = {1--12},
}

@article{pourahmadi_unit_2025,
	title = {Unit {Commitment} {Predictor} {With} a {Performance} {Guarantee}: {A} {Support} {Vector} {Machine} {Classifier}},
	volume = {40},
	issn = {1558-0679},
	shorttitle = {Unit {Commitment} {Predictor} {With} a {Performance} {Guarantee}},
	url = {https://ieeexplore.ieee.org/document/10530167},
	doi = {10.1109/TPWRS.2024.3400405},
	number = {1},
	urldate = {2025-07-24},
	journal = {IEEE Transactions on Power Systems},
	author = {Pourahmadi, Farzaneh and Kazempour, Jalal},
	month = jan,
	year = {2025},
	pages = {715--727},
}

@article{mohammadi_surrogate_2024,
	title = {Surrogate {Modeling} for {Solving} {OPF}: {A} {Review}},
	volume = {16},
	copyright = {http://creativecommons.org/licenses/by/3.0/},
	issn = {2071-1050},
	shorttitle = {Surrogate {Modeling} for {Solving} {OPF}},
	url = {https://www.mdpi.com/2071-1050/16/22/9851},
	doi = {10.3390/su16229851},
	language = {en},
	number = {22},
	urldate = {2025-07-24},
	journal = {Sustainability},
	publisher = {Multidisciplinary Digital Publishing Institute},
	author = {Mohammadi, Sina and Bui, Van-Hai and Su, Wencong and Wang, Bin},
	month = jan,
	year = {2024},
	note = {Number: 22},
	pages = {9851},
}

@article{bedi_deep_2019,
	title = {Deep learning framework to forecast electricity demand},
	volume = {238},
	issn = {0306-2619},
	url = {https://www.sciencedirect.com/science/article/pii/S0306261919301217},
	doi = {10.1016/j.apenergy.2019.01.113},
	urldate = {2025-07-23},
	journal = {Applied Energy},
	author = {Bedi, Jatin and Toshniwal, Durga},
	month = mar,
	year = {2019},
	pages = {1312--1326},
}

@article{nitsch_applying_2024,
	title = {Applying machine learning to electricity price forecasting in simulated energy market scenarios},
	volume = {12},
	issn = {2352-4847},
	url = {https://www.sciencedirect.com/science/article/pii/S2352484724007327},
	doi = {10.1016/j.egyr.2024.11.013},
	urldate = {2025-07-23},
	journal = {Energy Reports},
	author = {Nitsch, Felix and Schimeczek, Christoph and Bertsch, Valentin},
	month = dec,
	year = {2024},
	pages = {5268--5279},
}

@article{tschora_electricity_2022,
	title = {Electricity price forecasting on the day-ahead market using machine learning},
	volume = {313},
	issn = {0306-2619},
	url = {https://www.sciencedirect.com/science/article/pii/S0306261922002057},
	doi = {10.1016/j.apenergy.2022.118752},
	urldate = {2025-07-23},
	journal = {Applied Energy},
	author = {Tschora, Léonard and Pierre, Erwan and Plantevit, Marc and Robardet, Céline},
	month = may,
	year = {2022},
	pages = {118752},
}

@article{wogrin_lego_2022,
	title = {{LEGO}: {The} open-source {Low}-carbon {Expansion} {Generation} {Optimization} model},
	volume = {19},
	copyright = {All rights reserved},
	issn = {23527110},
	url = {https://doi.org/10.1016/j.softx.2022.101141},
	doi = {10.1016/j.softx.2022.101141},
	urldate = {2023-01-26},
	journal = {SoftwareX},
	author = {Wogrin, S. and Tejada-Arango, D.A. and Gaugl, R. and Klatzer, T. and Bachhiesl, U.},
	year = {2022},
}






\end{document}